%% file: RuleLangBench.tex
\newcommand{\arxiv}[1]{}
\newcommand{\oops}[1]{}
\newcommand{\iclp}[1]{#1}      
\newcommand{\noticlp}[1]{}     

\arxiv{
\documentclass[12pt]{article}
\usepackage{fullpage}
}

\documentclass[copyright,creativecommons]{eptcs}
\providecommand{\event}{ICLP 2023} 
\usepackage{iftex}
\ifpdf
  \usepackage[english]{babel}
\else
  \usepackage{breakurl}           
\fi

\renewcommand{\baselinestretch}{0.928}\small
\usepackage{float}

\usepackage{xspace}
\usepackage{alltt}
\usepackage[T1]{fontenc}

\usepackage[dvipsnames]{xcolor}
\definecolor{codegray}{rgb}{0.5,0.5,0.5}
\definecolor{codepurple}{rgb}{0.58,0,0.82}

\usepackage{amsmath,amssymb}
\usepackage{graphicx}
\usepackage{multirow}
\usepackage{multicol}
\usepackage{setspace} 

\usepackage{tikz}
\usepackage{pgfplots}
\usepackage{csvsimple}
\usepackage{filecontents}

\usepackage{pgfplotstable} 
\pgfplotsset{compat=1.15}

\usepackage{listings}

\lstdefinestyle{mystyle}{
    backgroundcolor=\color{white},   
    basicstyle=\footnotesize\ttfamily,
    commentstyle=\color{brown},
    keywordstyle=\color{blue}, 
    stringstyle=\color{codepurple},
    numberstyle=\tiny\color{codegray},
    numbers=none,
    breakatwhitespace=false,
    breaklines=false,                 
    captionpos=b,
    keepspaces=true,                 
    numbersep=5pt,
    showspaces=false,                
    showstringspaces=false,
    showtabs=false,                  
    tabsize=2
}
\newcommand{\Hex}[1]{\hspace{#1ex}}
\newcommand{\Vex}[1]{\vspace{#1ex}}

\newenvironment{code}{\Vex{0}\begin{alltt}\footnotesize}{\end{alltt}\Vex{0}}
\NewDocumentCommand{\co}{+m}{\mbox{\footnotesize\tt #1}} 
\newcommand\kw[1]{\textcolor{blue}{#1}} 
\newcommand{\mathify}[1]{\ifmmode{\mbox{$#1$}}\else\mbox{$#1$}\fi}
\newcommand\m[1]{$#1$} 
\newcommand\p[1]{\mathify{\it #1}} 

\newcommand\WHILE{\kw{while}\xspace}
\newcommand\INFER{\kw{infer}}
\newcommand\RULES{\kw{rules}}

\newcommand{\mysec}[1]{\Vex{-0}\section{#1\Vex{-0}}}
\newcommand{\mysubsec}[1]{\Vex{-0}\subsection{#1\Vex{-0}}}
\newcommand{\mypar}[1]{\Vex{-0.5}\paragraph{\bf #1.}}

\newcommand{\notes}[1]{}

\usepackage{longtable}

\usepackage{hyperref} 

\newcommand\fund{%
    This work was supported in part by NSF under grants 
    CCF-1954837, 
    CCF-1414078, and 
    IIS-1447549 
    and ONR under grants
    N00014-21-1-2719, 
    N00014-20-1-2751, and 
    N00014-15-1-2208. 
}

\begin{document}

\title{\m{\!\!\!}Benchmarking\,for\,Integrating\,Logic\,Rules\,with\,Everything\,Else\m{\!\!}}

\author{Yanhong A.\ Liu \quad\quad Scott D.\ Stoller \quad\quad
  Yi Tong \quad\quad K.\,Tuncay Tekle
  \institute{Computer Science Department, Stony Brook University, 
    Stony Brook, New York, USA}
  \email{\{liu,stoller,yittong,tuncay\}@cs.stonybrook.edu}}

\def\titlerunning{Benchmarking for Integrating Logic Rules with Everything
  Else}

\def\authorrunning{Y.\ A.\ Liu, S.\ D.\ Stoller, Y.\ Tong, and
  K.\ T.\ Tekle}

\maketitle

\begin{abstract}
Integrating logic rules with other language features is increasingly sought
after for advanced applications that require knowledge-base capabilities.
To address this demand, increasingly more languages and extensions for 
such integration have been developed.  How to evaluate such languages?

This paper describes a set of programming and performance benchmarks for
evaluating languages supporting integrated use of rules and other features,
and the results of evaluating such an integrated language together with
logic languages and languages not supporting logic rules.
\end{abstract}

\mysec{Introduction}

As knowledge-base capabilities are increasingly needed for advanced
applications, especially
applications that require sophisticated reasoning, logic rules are
increasingly needed together with other language features.  This leads to
increasingly more languages and extensions that provide such programming
support, for using rules through libraries
or directly as built-ins,
in many different ways.
%
%
%
To evaluate such language support for integrated uses of rules with other
language features, a set of benchmarking problems is needed.

Many benchmarking suites have been developed for evaluating logic languages
and rule engines, and even drastically more for imperative and other
languages.  In contrast, benchmarks and evaluation for languages that
support integrated use of rules together with other language features have
been lacking.

This paper describes a set of programming and performance benchmarks for
such evaluation, and the uses of these benchmarks in evaluating 
an integrated language together with logic languages and languages not
supporting logic rules.
The benchmarks are developed and organized into three sets:
\begin{description}

\item OpenRuleBench---
  for problems focused on using rules and evaluating
  rule systems, including all problems from
  OpenRuleBench~\cite{Lia+09open}, but modified and organized to use more
  appropriate language features for the benchmarking itself, especially
  imperative scripts for running and taking measurements.

\item RBAC---for problems that require (1) frequent imperative updates
  interleaved with expensive queries and (2) objects and classes with
  inheritance to organize system components, subsuming full Role-Based
  Access Control (RBAC)~\cite{ansi04role,LiuSto07RBAC-ONR}, with queries
  expressed in different ways of using rules and not using rules.

\item PA---for problems expressed using rules together with set queries and
  aggregate queries and recursive functions, from problems in program
  analysis (PA), because aggregate queries are particularly important for
  many database, data mining, and machine learning applications, but are
  not in OpenRuleBench~\cite{Lia+09open}.

\end{description}
The input data and workload for these benchmarks are designed for runs with
different problem scales---%
%
(1) large input data for queries using rules,
(2) large query results from inference using rules,
(3) large number of rules,
(4) frequent switches between using rules and using other features, and
(5) frequent invocations of inference using rules.
%

The evaluation is for benchmark problems written in (1) Alda, (2) XSB, (3)
Python, and (4) DistAlgo.
XSB~\cite{SagSW94xsb,\noticlp{swift2012xsb,}xsb22} is a top logic rule
engine for queries using rules, selected for reasons stated in
Section~\ref{sec-orb}.  Python~\cite{python23} is most used for its rich
features including high-level queries but lacks logic rules.
DistAlgo~\cite{\noticlp{Liu+12DistPL-OOPSLA,}Liu+17DistPL-TOPLAS,distalgo22github}
extends Python for distributed programming with higher-level and faster
queries but still lacks logic rules.  Alda~\cite{Liu+23RuleLangInteg-TPLP} is
an integrated language that supports logic rules as built-ins by building
on DistAlgo and employing XSB.
%
The evaluation yields the following main results:
\begin{itemize}

\item Queries using rules in Alda and XSB are similar, and are (1) simpler
  than using \co{while} loops in Python and DistAlgo, even when these loops
  use high-level set queries, and (2) drastically, asymptotically faster
  than \co{while} loops with high-level set queries in Python and DistAlgo.

\item Benchmarking using more generic imperative scripts in Alda is
  drastically simpler than using rules in XSB in OpenRuleBench; and
  integrated use of rules with other features for RBAC and PA problems
  yields simpler programs than not using rules or only using rules.

\item The performance of Alda programs is the accumulated performance of
  queries in XSB, other features in Python and DistAlgo, and interface
  between XSB and Python. Thus it is expected to increase accordingly when
  any part used is improved, 
  including reducing the current interface overhead to 1\% of it.
  %
  %
  
\item 
  Even with the current 
  interface overhead, Alda programs are faster
  or even drastically faster than half or more of the rule engines tested
  in OpenRuleBench~\cite{Lia+09open} for all but one benchmark, because the
  overhead is only linear in sizes of data and results and number of
  queries, and thanks to the XSB query performance.
\end{itemize}
Our analysis supports that similar results can be obtained by using other
efficient rule engines, including ASP systems, and other languages not
supporting logic rules, including lower-level languages. Our implementation
has been available by request and will be made public pending improved
documentation.

\mysec{Language}
\label{sec-lang}

XSB and Python are well established, and DistAlgo is subsumed by Alda.  So
we
first introduce Alda, an integrated 
language that supports rules with all
of sets, functions, updates, and objects, all as built-ins, seamlessly
integrated without extra interface code.
Figure~\ref{fig-prog} shows an example program in Alda.
\begin{figure}[t]
\centering
\begin{lstlisting}[numbers=left, basicstyle=\scriptsize\ttfamily]
class CoreRBAC:                  # class for Core RBAC component/object
  def setup():                   # method to set up the object, with no arguments
    self.USERS, self.ROLES, self.UR := {},{},{}
                                 # set users, roles, user-role pairs to empty sets
  def AddRole(role):             # method to add a role
    ROLES.add(role)              # add the role to ROLES
  def AssignedUsers(role):       # method to return assigned users of a role
    return {u: u in USERS | (u,role) in UR}  # return set of users having the role
\end{lstlisting}
%
\Vex{-3}
\begin{lstlisting}[basicstyle=\scriptsize\ttfamily]
  ...
\end{lstlisting}
\Vex{-2}
\begin{lstlisting}[numbers=left,firstnumber=9, basicstyle=\scriptsize\ttfamily]
class HierRBAC extends CoreRBAC: # Hierarchical RBAC extending Core RBAC
  def setup():
    super().setup()              # call setup of CoreRBAC, to set sets as in there
    self.RH := {}                # set ascendant-descendant role pairs to empty set
  def AddInheritance(a,d):       # to add inherit. of an ascendant by a descendant
    RH.add((a,d))                # add pair (a,d) to RH
  rules trans_rs:                # rule set defining transitive closure
    path(x,y) if edge(x,y)       # path holds for (x,y) if edge holds for (x,y)
    path(x,y) if edge(x,z), path(z,y)  # ... if edge holds for (x,z) and for (z,y)
  def transRH():                 # to return transitive RH and reflexive role pairs
    return infer(path, edge=RH, rules=trans_rs) + {(r,r): r in ROLES}
  def AuthorizedUsers(role):     # to return users having a role transitively
    return {u: u in USERS, r in ROLES | (u,r) in UR and (r,role) in transRH()}
\end{lstlisting}
\Vex{-3}
\begin{lstlisting}[basicstyle=\scriptsize\ttfamily]
  ...
\end{lstlisting}
\Vex{-2}
\begin{lstlisting}[numbers=left,firstnumber=22, basicstyle=\scriptsize\ttfamily]
h = new(HierRBAC, [])            # create HierRBAC object h, with no args to setup
h.AddRole('chair')               # call AddRole of h with role 'chair'
\end{lstlisting}
\Vex{-3}
\begin{lstlisting}[basicstyle=\scriptsize\ttfamily]
...
\end{lstlisting}
\Vex{-2}
\begin{lstlisting}[numbers=left,firstnumber=24, basicstyle=\scriptsize\ttfamily]
h.AuthorizedUsers('chair')       # call AuthorizedUsers of h with role `chair'
\end{lstlisting}
\Vex{-3}
\begin{lstlisting}[basicstyle=\scriptsize\ttfamily]
...
\end{lstlisting}
\Vex{-2}
\caption{An example program in Alda, for Role-Based Access Control (RBAC), demonstrating logic rules used with sets, functions, updates, and objects.}
\label{fig-prog}
\end{figure}

Rules are defined using rule sets (e.g. lines 15--17, with rule set name
\co{trans\_rs}), and queried using calls to an inference function,
\co{infer} (e.g. on line 19,
using rule set \co{trans\_rs}).
The two rules in \co{trans\_rs} (lines 16--17) define predicate \co{path}
using predicate \co{edge}, where the first rule is the base case, and the
second rule is the recursive case.
The call to \co{infer} (line 19) returns the result of querying \co{path},
i.e., the set of pairs for which \co{path} holds, given that \co{edge}
equals \co{RH}, i.e., \co{edge} holds for the set of pairs in \co{RH}.
%

In a rule set, predicates not in any conclusion are called base predicates;
the other predicates are called derived predicates.
The key idea is that predicates are simply set-valued variables and vice
versa.  So predicates can be defined and used directly as other variables,
in any scope---global, object field, or local to a rule set.  The only
exception is that derived predicates can only be updated by \co{infer}, 
and are automatically maintained
by an implicit call to \co{infer} when any non-local base predicates are
updated.  The exception ensures the declarative semantics of rules.

All other features---class, inheritance, method (function and procedure),
update, and set query---are the same as in object-oriented languages.  We
mostly use Python syntax except for a few conventions from Java (\co{extends} for inheritance, \co{new}
for object creation, and omission of
\co{self} when there is no ambiguity) for ease of reading, and a few ideal
syntax (\co{:=} for assignment, \co{\{\}} for empty set, and \co{+} for set
union), and pattern matching in queries (in examples explained later).

For the example in Figure~\ref{fig-prog} but without using rules, computing
the transitive closure (i.e., result of \co{infer} in function
\co{transRH}) in a variable \co{T}, by adding a parameter \co{E} to
function \co{transRH} and passing the value of \co{RH} to \co{E} when
calling \co{transRH}, can use a \co{while} loop in Python, as used
for~\cite{LiuSto07RBAC-ONR}:
\begin{lstlisting}
  T = E.copy()
  W = {(x,y) for (x,z) in T for (z2,y) in E if z2==z} - T
  while W:
    T.add(W.pop())
    W = {(x,y) for (x,z) in T for (z2,y) in E if z2==z} - T
\end{lstlisting}
or a \co{while} loop with higher-level set queries in DistAlgo:
\begin{lstlisting}
  T := E.copy()
  while some (x,z) in T, (z,y) in E | (x,y) not in T:
    T.add((x,y))
\end{lstlisting}
Using rules is clearly simpler, both conceptually and in the amount of code.

Alda is implemented by extending the DistAlgo
compiler~\cite{\noticlp{Liu+12DistPL-OOPSLA,}Liu+17DistPL-TOPLAS,distalgo22github}
and invoking XSB~\cite{SagSW94xsb,\noticlp{swift2012xsb,}xsb22} for
inference.  DistAlgo
is compiled to Python.
Function \co{infer} translates data from Python to Prolog facts,
translates results back, and invokes XSB in between using command line and
file passing.  The obvious overhead of this external interface can be
removed with an in-memory interface between Python to XSB, which is
actively being developed by the XSB team.\footnote{A version working for
  Unix, not yet Windows, has been released, and passing lists of length 100
  million in memory took about 30 nanoseconds per element~\cite[release
  notes]{xsb22}.  So even for the largest data in our experiments, of size
  a few millions, it would take 0.1-0.2 seconds to pass in memory, instead
  of 10-20 seconds with the current external interface.}  However, this
has not affected Alda programs from having generally good performance,
thanks to the XSB query performance.

\input{bench}

\input{experiment}

\mysec{Related work and conclusion}

Many benchmarking suites have been developed for evaluating the performance
of queries in logic languages and rule engines, including the carefully
constructed OpenRuleBench for comprehensively evaluating a diverse set of problems in a wide range of systems~\cite{Lia+09open}.
In particular, Prolog has had benchmarks for comparing performance of
different implementations~\cite{cmu85prologbench,unipr01clpbench}.
Some are focused on a special class of problems, e.g., interpreters written
in Prolog~\cite{korner2020performance}.  Some are for evaluating the
performance of a particular implementation, e.g., SWI
Prolog~\cite{swi22bench}.
There are also works that evaluate queries in more general rule engines and
database systems, e.g., the LUBM benchmark~\cite{guo2005lubm} and its
extensions~\cite{ma2006towards,singh2020owl2bench} for OWL
on a range of systems,
and evaluations including SQL with
rules~\cite{john2019new,brass2019performance}.
There are also works focused on evaluating interfaces, e.g., Java Prolog
Interface~\cite{llanes2022java}.
These works study only queries.

Even drastically more benchmarking suites were developed for evaluating
other systems.
Works range from systematic studies, e.g.,~\cite{lewis1985evolution,gray93benchmark,nambiar2009performance},
to a large variety of specific benchmarks, e.g., the SPEC benchmarks for computer systems in general~\cite{nambiar2009performance}, the TPC benchmarks for transaction processing~\cite{nambiar2010transaction}, the LINPACK benchmarks~\cite{dongarra2003linpack} on solving linear equations, and many more. 
These benchmarks exercise updates and many other language features but not logic rules.

In contrast, our work develops benchmarks that exercise integrated use of
rules together with other language features.  We improve OpenRuleBench
benchmarks with significantly simplified benchmarking code, and design new
benchmarks that exercise tightly integrated uses of different features on
problems that test different problem scales in different ways. We also
compare different ways of using rules vs.\ not using rules\noticlp{, as
  well as performance}.

In conclusion, this work presents a set of programming and performance
benchmarks for evaluating languages supporting integrated use of rules and
other features, and the results of using these benchmarks in an evaluation.
Future work can perform evaluations with additional languages and systems,
especially efficient ASP systems such as Clingo~\cite{gebser2019multi}.
\noticlp{,as well as evaluation of more metrics such as memory usage.}

\mypar{Acknowledgments}
We thank Bo Lin for his robust DistAlgo compiler implementation, and his
initial help in implementing the Alda compiler, and we thank David S.\
Warren for his initial help in implementing an interface to XSB.  We also
thank Thang Bui for additional applications in program analysis and
optimization, and students in undergraduate and graduate courses for using
Alda and its earlier versions, called DA-rules.
\fund

{
\renewcommand{\baselinestretch}{-.1}\small
\def\usebib{
\def\bibdir{../../../bib}   
{
\bibliography{\bibdir/strings,\bibdir/liu,\bibdir/IC,\bibdir/PT,\bibdir/PA,\bibdir/Lang,\bibdir/Algo,\bibdir/DB,\bibdir/AI,\bibdir/Sec,\bibdir/Sys,\bibdir/Perform,\bibdir/SE,\bibdir/Vis,\bibdir/misc,\bibdir/crossref} 
\bibliographystyle{eptcs}
}
}
\usebib
}

\end{document}

%% file: bench.tex
\mysec{Programming and performance benchmarks}
\label{sec-bench}

\iclp{This section presents three sets of benchmarks that we developed.
They are written to express each given problem in the most direct way 
we can think of.  This helps make the benchmarks clear and easy to read.
This also avoids being unfair to any particular language or system being
compared with.
}%
\noticlp{
We have used Alda for a 
variety of well-known problems where rules can
be used for both ease of programming and performance of execution.
We describe a set of benchmarks for programming and performance evaluation.
One can see that even for problems that were previously focused on for
using rules, it becomes much easier to program using an integrated language
like Alda.

We developed three sets of benchmarks, from
OpenRuleBench~\cite{Lia+09open}, RBAC~\cite{ansi04role,LiuSto07RBAC-ONR},
and program analysis.
OpenRuleBench benchmarks show the wide range of application problems
previously developed using different kinds of rule systems.
RBAC benchmarks show the use of rules in an application that requires all
of sets, functions, updates, and objects and classes, and show different
ways of using rules.
Program analysis benchmarks demonstrate seamlessly integrated use of rules
with also aggregate queries and recursive functions; we also contrast with
using aggregate queries in rule languages, which are not used in any
benchmark in OpenRuleBench.\oops{ The latter two sets are described in the
  Supplementary Material.}
}

\mysubsec{OpenRuleBench---a wide variety of rule-based applications}
\label{sec-orb}

OpenRuleBench~\cite{Lia+09open} contains a wide variety of database,
knowledge base, and semantic web application problems, written using rules
in 11 well-known rule systems from 5 different categories, as well as large
data sets and a large number of test scripts for running and measuring the
performance.  Among 14 benchmarks described in~\cite{Lia+09open}, we
consider all except for one that tests interfaces of rule systems with
databases (which is a non-issue for Python and its extensions DistAlgo and
Alda, because Python has standard and widely-used database interfaces).

Table~\ref{tab-rulebench} summarizes the benchmarks.  We compare with the
benchmark programs in XSB, for three reasons: (1) XSB has been the most
advanced rule system supporting well-founded semantics for non-stratified
negation and tabling techniques for efficient query evaluation, and has
been actively developed for over three decades, to this day;
(2) among all systems reported in~\cite{Lia+09open}, XSB
was one of the fastest, if not the fastest, and the single most consistent across
all benchmarks; and (3) among all measurements reported, only XSB,
OntoBroker, and DLV could run all benchmarks, but OntoBroker went bankrupt,
and measurements for DLV were almost all slower, often by orders of
magnitude.

We easily translated all 13 benchmarks into Alda, automatically for all
except for three cases where the original rules used features beyond
Datalog, which became two cases after we added support for negation.  In
all cases, it was straightforward to express the desired functionality in
Alda, producing a program that is very similar to XSB or even simpler.
Additionally, the code for reading data, running tests, timing, and writing
results is drastically simpler in Alda than in XSB, because Alda extends
Python.  The three exception cases and additional findings are described
below.

\begin{table}[t]
  \small
  \centering
\begin{tabular}{@{~}l@{~}|@{~}p{52ex}@{~}|@{~}r@{~}|@{~}r@{~}}
  Name    & Description & Prog size & Data size \\\noticlp{\hline}\iclp{\cline{1-4}}
  \iclp{\noalign{\vskip .4ex}}
  \noticlp{\hline}\iclp{\cline{1-4}}
  Join1   & non-recursive tree of binary joins as inference rules
                        &  225 & * \\\noticlp{\hline}\iclp{\cline{1-4}}
  Join2   & join from IRIS system producing large intermediate result
                        &   41 & * \\\noticlp{\hline}\iclp{\cline{1-4}}
  JoinDup & join of separate results of five copies of Join1
                        &  163 & * \\\noticlp{\hline}\iclp{\cline{1-4}}
  LUBM    & university database adapted from LUBM benchmark
                        &  377 & * \\\noticlp{\hline}\iclp{\cline{1-4}}
  Mondial & geographical database derived from CIA Factbook
                        &   36 & 59,733\\\noticlp{\hline}\iclp{\cline{1-4}}
  DBLP    & well-known bibliography database on the Web
                        &   20 & 2,437,867\\\noticlp{\hline}\iclp{\cline{1-4}}
  \iclp{\noalign{\vskip .4ex}}
  \noticlp{\hline}\iclp{\cline{1-4}}
  TC      & classical transitive closure of a binary relation
                        &   75 & * \\\noticlp{\hline}\iclp{\cline{1-4}}
  SG      & well-known same-generation siblings problem
                        &   90 & * \\\noticlp{\hline}\iclp{\cline{1-4}}
  WordNet & natural language processing queries based on WordNet
                        &  298 & 465,703\\\noticlp{\hline}\iclp{\cline{1-4}}
  Wine    & well-known OWL wine ontology as rules
                        & 1103 & 654 \\\noticlp{\hline}\iclp{\cline{1-4}}
  \iclp{\noalign{\vskip .4ex}}
  \noticlp{\hline}\iclp{\cline{1-4}}
  ModSG   & modified SG to exclude ancestor-descendant relationships
                        &   38 & * \\\noticlp{\hline}\iclp{\cline{1-4}}
  Win     & well-known win-not-win game with non-stratified negation
                        &   24 & * \\\noticlp{\hline}\iclp{\cline{1-4}}
  MagicSet & non-stratified rules from magic-set transformation
                        &   34 & * \\\noticlp{\hline}\iclp{\cline{1-4}}
  \iclp{\noalign{\vskip .4ex}}
  \noticlp{\hline}\iclp{\cline{1-4}}
\end{tabular}\Vex{.5}\\
\def\taborbcaption{Benchmarks from OpenRuleBench.}
\begin{tabular}{@{}p{100ex}}
  The three groups (of 6, 4, 3) in order are called 
  large join tests,
  Datalog recursion, and default negation, respectively.
  Prog size is the XSB program size 
  in lines of code without comments and empty lines.
  Data size is the input data size in number of facts; 
  * means that scripts are used to generate input data of desired sizes.\Vex{-1}
\end{tabular}
\caption{\taborbcaption}
\label{tab-rulebench}
\end{table}

\mypar{Result set}
In most logic languages, including Prolog and many variants, a query
returns only the first result that matches the query.  To return the set of
all results, some well-known tricks are used.  The LUBM benchmark
includes the following extra rules to return all answers of \co{query9\_1}:
\begin{code}
  query9 :- query9_1(X,Y,Z), fail.
  query9 :- writeln('========query9.======').
\end{code}
The first rule first queries \co{query9\_1} to find an answer (a triple of
values for \co{X},\co{Y},\co{Z}) and then uses \co{fail} to trick the
inference into thinking that it failed to find an answer and so continuing
to search for an answer; and it does this repeatedly, until \co{query9\_1}
does fail to find an answer after exhausting all answers.  The second rule
is necessary, even if with an empty right side, to trick the inference into
thinking that it succeeded, because the first rule always ends in failing;
this is so that the execution can continue to do the remaining work instead
of stopping from failing.

In fact, this trick is used in all benchmarks, but other uses are buried
inside the code for running, timing, etc., specialized for each benchmark,
not as part of the rules for the application logic.

In Alda, such rules and tricks are never needed.  A call to \co{\INFER}
with query \co{query9\_1} returns the set of all query results as desired.
\noticlp{
If \co{query9\_1} is a non-local predicate, then the set value of
\co{query9\_1} can be used directly, and no explicit call to \co{\INFER} is
needed.}
\arxiv{In case only one result is wanted, a special function for taking any
  one value can be applied to the result set of calling \co{\INFER} or the
  non-local predicate; an optimized implementation can then search for only
  the first result.}

\mypar{Function symbols}
Logic rules may use function symbols to form structured data
that
can be used as
arguments to predicates.  Uses of function symbols can be translated away.
The Mondial benchmark uses a function symbol \co{prov} in several intermediate conclusions
and hypotheses of the form \co{isa(prov(Y,X),provi)} or \co{att(prov(Y,X),number,A)}.
They can simply be translated to 
\co{isa('prov',Y,X,provi)} and\linebreak \co{att('prov',Y,X,number,A)},
respectively.

\mypar{Negation}
Logic languages may use negation applied to hypotheses in rules.
Most logic languages only support stratified negation, where there is no
negation involved in cyclic dependencies among predicates.  Such negation
can be done by set differences.  The ModSG benchmark has such a negation,
as follows, where \co{sg} is defined by the rules in the SG benchmark, and
\co{nonsg} is defined by two new rules.
\begin{code}
  sg2(X,Y) :- sg(X,Y), not nonsg(X,Y).
\end{code}
In Alda, this can be written as 
\begin{code}
  sg2 = \INFER(sg, \RULES=sg\_rs) - \INFER(nonsg, \RULES=nonsg\_rs)
\end{code}
where \co{sg\_rs} and \co{nonsg\_rs} are the rule sets
defining \co{sg} and \co{nonsg}, respectively.
\noticlp{,
 and all base predicates are non-local.}

Alda also supports negation in rules. Its current implementation translates
negation to tabled negation \co{tnot} in XSB.  This handles even
non-stratified negation by computing well-founded semantics using
XSB~\cite{CheWar96}, contrasting Prolog's negation as failure.  The Win and
MagicSet benchmarks have non-stratified negation.
Both of them, as well as ModSG, can be expressed directly in Alda rule sets
by using \co{not} for negation.

\mypar{Benchmarking and organization} 
In OpenRuleBench benchmarks, even though the rules to be benchmarked are
declarative and succinct, the benchmarking code for reading input, running
tests, timing, and writing results are generally much larger.  For example,
the Join1 benchmark has 4 small rules and 9 small queries similar in size
to those in the transitive closure example, plus a manually added tabling
directive for optimization.  However, for each of the 9 queries, 19 more
lines for an import and two much larger rules are used to do the reading,
running, timing, and writing.

In general, because benchmarking executes a bundle of commands, scripting
those directly is simplest.  Furthermore, organizing benchmarking code
using procedures, objects, etc., allows easy reuse without duplicated code.
These features are much better supported in languages like Python than rule
systems, for both ease of programming and performance,

In fact, OpenRuleBench uses a large number of many different files, 
in several languages (language of the system being tested, XSB, shell script, 
Python, makefile) for such scripting.  
For example for Join1, the 4 rules, tabling directive, and benchmarking code
are also duplicated in each of the 9 XSB files, one for each query; 
a 46-line shell script and a 9-line makefile are also used.

In contrast, our benchmarking code is in Alda, which uses Python functions
for scripting.\oops{ A same 30-line Alda program is used to benchmark any
  of the benchmarks, and a 16-line Alda program reads any file of facts and
  writes a pickled file}
A single 45-line Alda program is used 
for timing any of the benchmarks, and 
for pickling (i.e., object serialization in Python, for fast data 
reading after the first reading) and timing of pickling.

%
%
\mypar{Aggregation}
Despite the wide variety of benchmarks in OpenRuleBench, 
no benchmark uses aggregate queries.  Aggregate queries are essential for
many database, data mining, and machine learning applications.  We discuss
them and compare with aggregate queries in a rule language like XSB in
Section~\ref{sec-pa}.

\mysubsec{RBAC---rules with objects, updates, and set queries}
\label{sec-rbac}
%

\noticlp{As discussed in Section~\ref{sec-obj}, a complex and inefficient
\co{\WHILE} loop was used in \cite{LiuSto07RBAC-ONR} to program the
transitive role hierarchy, but as discussed in Section~\ref{sec-prob}, an
efficient algorithm with appropriate data structures and updates would be
drastically even more complex.}
\iclp{The ANSI standard for Role-Based Access Control (RBAC~\cite{ferraiolo01proposed,ansi04role} involves many sets and query and update functions in a total of 9 components. To program the transitive role hierarchy at a high level, a complex and inefficient
\co{\WHILE} loop was used before~\cite{LiuSto07RBAC-ONR}, because an
efficient algorithm would be drastically even more complex.}

With support for rules, the entire RBAC standard is easily written in Alda,
similar as in Python~\cite{LiuSto07RBAC-ONR}, except with rules for
computing the transitive role hierarchy, as \noticlp{described in
Section~\ref{sec-obj}}\iclp{shown in Figure~\ref{fig-prog}},\arxiv{ and with simpler set queries and omission of
  \co{self},}\noticlp{ despite complex class inheritance relationships,} yielding a
simpler yet more efficient program.

Below, we specify \noticlp{different}\iclp{more} ways of using rules to compute the transitive
role hierarchy and\noticlp{ the} function
\co{AuthorizedUsers(role)} in Figure~\ref{fig-prog}.
All these ways are declarative and differ in size by only 1-2 lines.
Table~\ref{tab-rbac} summarizes the benchmarks for RBAC that include all
RBAC classes with their inheritance relationships and perform update
operations and these query functions in different ways.
%

In particular, in the first way below, a field, \co{transRH}, is used and
maintained automatically; it avoids calling \co{transRH()} repeatedly, as
desired in the RBAC standard, and it does so without the extra maintenance
code in the RBAC standard for handling updates.

\begin{table}[t]
  \small
  \centering
\begin{tabular}{@{~}l@{~}|@{~}p{61ex}}
  Name    & Features used for computing transitive role hierarchy
  \\\noticlp{\hline}\iclp{\cline{1-2}}

  RBACnonloc & rule set \co{transRH\_rs} with 
             implicit \co{\INFER}, without \co{transRH()}, Sec.\,\ref{sec-rbac}
  \\\noticlp{\hline}\iclp{\cline{1-2}}
  RBACallloc & rule set \co{trans\_role\_rs} 
             and \co{transRH()} that has only \co{\INFER}, Sec.\,\ref{sec-rbac}
  \\\noticlp{\hline}\iclp{\cline{1-2}}
  RBACunion & rule set \co{trans\_rs} and 
             \co{transRH()} that has \co{\INFER} and union, Sec.\,\ref{sec-\noticlp{obj}\iclp{lang}}
  \\\noticlp{\hline}\iclp{\cline{1-2}}
  RBACda    & \co{\WHILE} loop and 
             high-level set queries in DistAlgo, Sec.\,\ref{sec-\noticlp{pred}\iclp{lang}}
  \\\noticlp{\hline}\iclp{\cline{1-2}}
  RBACpy    & \co{\WHILE} loop and
             high-level set comprehensions in Python\noticlp{~\cite{LiuSto07RBAC-ONR}}\iclp{, Sec.\,\ref{sec-lang}}
  \\\noticlp{\hline}\iclp{\cline{1-2}}
\end{tabular}\Vex{.5}\\
\def\tabrbaccaption{Benchmarks for RBAC updates and queries.}
\begin{tabular}{@{}p{100ex}}
  Each benchmark performs a combination of updates to sets and relations \co{USERS},
  \co{ROLES}, \co{UR}, and \co{RH} and queries with function
  \co{AuthorizedUsers(role)}, where the transitive role hierarchy is computed
  with a different way of using rules, or not using rules.
  In \co{AuthorizedUsers(role)} of all five programs, the call to
  \co{transRH()}, or reference to field \co{transRH}, is lifted out of the
  set query, by assigning its value to a local variable and using that
  variable in the query.\Vex{-2}
\end{tabular}
\caption{\tabrbaccaption}
\label{tab-rbac}
\end{table}

\mypar{Rules with only non-local predicates}
\label{sec-nonlocal}
Using rules with only non-local predicates, one can add a field \co{transRH},
and use \co{transRH} in place of calls to \co{transRH()}, e.g., in function
\co{AuthorizedUsers(role)},
and use the following rule set instead of \co{trans\_rs} in
class \co{HierRBAC}:\footnote{The first rule could actually be omitted,
  because
  the second argument of \co{RH} is always in \co{ROLES} and thus the
  second rule when joining \co{RH} with reflexive pairs in \co{transRH}
  from the third rule subsumes the first rule.}
\begin{lstlisting}
  rules transRH_rs:  # no need to use infer explicitly
    transRH(x,y) if RH(x,y)
    transRH(x,y) if RH(x,z), transRH(z,y)
    transRH(x,x) if ROLES(x)
\end{lstlisting}
Field \co{transRH} is automatically maintained at updates to \co{RH} and
\co{ROLES}
by implicit calls to \co{\INFER}; %
no explicit calls to \co{\INFER} are needed.  
This eliminates the need of function \co{transRH()} and repeated expensive
calls to it even when its result is not changed most of the time.
Overall, this simplifies the program, ensures correctness, and improves
efficiency.

\mypar{Rules with only local predicates}
\label{sec-appl-local}
Using rules with only local predicates, \co{\INFER} must be called
explicitly.
One can simply use the function \co{transRH()} in Figure~\ref{fig-prog},
which calls \co{\INFER} using rule set \co{trans\_rs} in the running
example and then unions with reflexive role pairs.  Alternatively, one can
use the rules in \co{trans\_rs} plus a rule that uses a local predicate
\co{role} to infer reflexive role pairs:
\begin{lstlisting}
  rules trans_role_rs:  # as trans_rs plus the added last rule
    path(x,y) if edge(x,y)
    path(x,y) if edge(x,z), path(z,y)
    path(x,x) if role(x)               
  def transRH():        # use infer only, pass in also ROLES
    return infer(path, edge=RH, role=ROLES, rules=trans_role_rs)
\end{lstlisting}
Both ways show the ease of using rules by simply calling \co{\INFER}.
%
Despite possible inefficiency in some cases, 
using only local predicates has the advantage of full reusability of rules 
and full control of calls to \co{\INFER}.

\noticlp{\mypar{Rules with both local and non-local predicates}
\label{sec-appl-mixed}
One can also use rules with a combination of local and non-local
predicates, e.g., the same rules as in \co{trans\_role\_rs} but with field \co{ROLES} in
place of the local \co{role},
removing the need for \co{\INFER} to pass in \co{ROLES}.
Any other combination 
can also be used.
Different combinations give different controls to \co{\INFER} to pass in and
out appropriate sets. 

Of course,
non-recursive set queries, such as \co{AuthorizedUsers(role)} can also be
expressed using rules, and use any combination of local and non-local
predicates.
}

\mysubsec{Program analysis---rules with aggregate queries and recursive
  functions}
\label{sec-pa}
%

We designed a benchmark for program analysis (PA) problems,
especially to show integrated use of rules with aggregate queries and
recursive functions.
Aggregate queries help quantify and characterize the analysis results,
and recursive functions help do these on recursive structures.
We describe programs written in Alda and then in XSB.

The benchmark is for analysis of class hierarchy of Python programs.  It
uses logic rules to extract class names and construct the class extension
relation;
aggregate queries and set queries to characterize the results and find
special cases of interest;
recursive functions as well as aggregate and set queries to analyze the
special cases;
and more logic rules, functions, and set and aggregate queries to further
analyze the special cases.

Table~\ref{tab-PA} summarizes different parts of this benchmark, called PA.
A variant, called PAopt, is the same as PA except that, in
the recursive rule for defining transitive descendant relationship, the two
hypotheses are reversed, following previously studied
optimizations~\cite{LiuSto09Rules-TOPLAS,TekLiu10RuleQuery-PPDP}.
%
\begin{table}[t]
  \small
  \centering
\begin{tabular}{@{~}l@{}@{~}l@{~}|@{~}p{29ex}@{~}|@{~}p{47ex}@{~}}
  \Hex{0}Part\Hex{-4}  & & Analysis & Features used \\\noticlp{\hline}\iclp{\cline{1-4}}
  1 & Ext  & classes, 
             extension relation 
                  & rules (recursive if refined name analysis is used)
  \\\noticlp{\hline}\iclp{\cline{1-4}}
  2 & Stat & statistics, roots
                  & aggregate and set queries
  \\\noticlp{\hline}\iclp{\cline{1-4}}
  3 & Hgt & max height, 
            roots of max height 
                  & recursive functions, 
                    aggregate and set queries
  \\\noticlp{\hline}\iclp{\cline{1-4}}
  4 & Desc & max desc, 
             roots of max desc
                  & recursive rules, functions, 
                    aggregate and set queries
  \\\noticlp{\hline}\iclp{\cline{1-4}}
\end{tabular}\Vex{.5}\\
\def\tabpacaption{  Benchmark PA for program analysis, integrating 
  different kinds\noticlp{ of analysis problems}.
}
\begin{tabular}{@{}p{100ex}}
  In Parts 1 and 4 that use rules,
  not using rules (esp.\ for recursive analysis, with tabling) 
  would be drastically worse
  (i.e., harder to program and less efficient).
  In Parts 2-4 that use aggregate and set queries, using
  rules or recursive functions would be clearly worse.
  In Parts 3 and 4 that use functions, not using functions (with tabling, also
  called caching) would be much worse.\Vex{-1}
\end{tabular}
\caption{\tabpacaption}
\label{tab-PA}
\end{table}

Because the focus is on evaluating the integrated use of different
features, each part that uses a single feature, such as rules, is
designed to be small.  Compared with making each part larger, which
exercises individual features more, this design highlights 
the overhead of
connecting different parts, in terms of both ease of use and efficiency of
execution. 

The benchmark program takes as input 
the abstract syntax tree (AST) of a Python program 
(a module or an entire package), represented as a set of facts.
Each AST node of type \p{T} with \p{k} children corresponds to 
a fact for predicate \p{T}
with \p{k}+1 arguments: id of the node, and
ids of the \p{k} children. 
Lists are represented using \co{Member} facts.  A
\co{Member(\p{lst},\p{elem},\p{i})} fact denotes that list \p{lst} has
element \p{elem} at the \p{i}th position.

\mypar{
Part 1: Classes and class extension relation} 
This part examines all \co{ClassDef} nodes in the AST.  A \co{ClassDef}
node has 5 children: class name, list of base-class expressions, and three
nodes not used for this analysis.
The following rules can be used to find all defined class names and build
a class extension relation using base-class expressions that are
\co{Name} nodes.  A \co{Name} node has two children: name and
context.
\begin{lstlisting}
  rules class_extends_rs:
    defined(c) if ClassDef(_,c,_, _,_,_)
    extending(c,b) if ClassDef(_,c,baselist, _,_,_),
                      Member(baselist,base,_), Name(base,b,_)
\end{lstlisting}
For a dynamic language like Python, analysis involving names can be refined
in many ways to give more precise results,
e.g.,~\cite{Gor+10Alias-DLS}.
We do not do those here, but Datalog rules are particularly good for
such analysis of bindings and aliases for names,
e.g.,~\cite{smara15pointer}.

\mypar{
Part 2: Characterizing results and finding special cases}
This part computes statistics for defined classes and the class extension
relation
and finds %
root classes (class with subclass but not super class).
%
These use aggregate queries and set queries\iclp{, where \co{(\_,c)} and \co{(=c,\_)} are tuple patterns, \co{\_} matches anything, \co{c} is bound to the matched value, and \co{=c} matches the value of \co{c}}.
\begin{lstlisting}
  num_defined := count(defined)
  num_extending := count(extending)
  avg_extending := num_extending/num_defined
  roots := {c: (_,c) in extending, not some (=c,_) in extending}
\end{lstlisting}
Similar queries can compute many other statistics and cases: maximum number
of classes that any class extends, leaf classes, histograms, etc.

\mypar{Part 3: Analysis of special cases}
This part computes the maximum height of the extension relation, which is the
maximum height of the root classes, and finds root classes of the maximum
height.
These use a recursive function as well as aggregate and set queries.
\begin{lstlisting}
  def height(c):
    return 0 if not some (_,=c) in extending
           else 1 + max{height(d): (d,=c) in extending}
  max_height := max{height(r): r in roots}
  roots_max_height := {r: r in roots, height(r) = max_height}
\end{lstlisting}
For efficiency when a subclass can extend multiple base classes, caching of
results of function calls is used.
In Python, one can simply add \co{import functools} to import module
\co{functools}, and add \co{@functools.cache} just above the definition of
\co{height} to 
cache the results of that function.

\mypar{Part 4: Further analysis of special cases}
This part computes the maximum number of descendant classes following the
extension relation from a root class, and finds root classes of the maximum
number.
Recursive functions and aggregate queries similar to finding maximum height
do not suffice here,
due to shared subclasses that may be at any depth.
Instead, the following rules can infer all \co{desc(c,r)} facts where class
\co{c} is a descendant following the extension relation from root class
\co{r}, and aggregate and set queries with function \co{num\_desc} then
compute the desired results.
\begin{lstlisting}
  rules desc_rs:
    desc(c,r) if roots(r), extending(c,r)
    desc(c,r) if desc(b,r), extending(c,b) 
\end{lstlisting}\Vex{-2}
\begin{lstlisting}
  def num_desc(r):
    return count{c: (c,=r) in desc}
  max_desc := max{num_desc(r): r in roots}
  roots_max_desc := {r: r in roots, num_desc(r) = max_desc}
\end{lstlisting}
For efficiency of the last query, caching is also used for function
\co{num\_desc}.  If the last query is omitted, function
\co{num\_desc} can also be inlined in the \co{max\_desc} query.

\mypar{Comparing with aggregate queries and functions in rule languages}
While rules in Alda correspond directly to rules in rule languages,
  expressing aggregate queries and functions using rules require
  translations that formulate computations as hypotheses and introduce
  additional variables to relate these hypotheses.

Aggregate queries are used extensively in database 
and machine learning applications, and are essential for analyzing large data
or uncertain information.  
These queries are easy to express directly in database languages and
scripting languages, but are less so in rule languages like Prolog; most
rule languages also do not support general aggregation with recursion due
to their subtle semantics~\cite{LiuSto22RuleAgg-JLC}.  For example, the
simple query \co{num\_defined := count(defined)} in Alda, when written in
XSB, becomes:
\begin{code}
  num_defined(N) :- setof(C, defined(C), S), length(S, N).
\end{code}

Recursive functions are used extensively in list and tree processing and in
solving divide-and-conquer problems.  They are natural for computing
certain information about parse trees, nested scopes,
etc.
However, in rule languages, they are expressed in a way that mixes function
arguments and return values, and require sophisticated mode analysis to
differentiate arguments from returns.  For example, the \co{height} query,
when written in XSB, becomes:
\begin{code}
  height(C,0) :- not(extending(_,C)).
  height(C,H) :- findall(H1, (extending(D,C), height(D,H1)), L),
                 max_list(L,H2), H is H2+1.
\end{code}

%% file: experiment.tex
\mysec{Experimental results}
\label{app-expe}

\newcommand{\myheading}[1]{\mypar{#1}}

We present results about running times, program sizes, and data sizes.
%
All measurements were taken on
a machine with an Intel Xeon X5690 3.47 GHz CPU, 94 GB RAM,
running 64-bit Ubuntu 16.04.7,
Python 3.9.9, and XSB 4.0.0.
%
For each experiment, the reported running times are CPU times averaged over
10 runs.  Garbage collection in Python was disabled for smoother running
times when calling XSB.
Program sizes are numbers of lines excluding comments and empty lines.
Data sizes are number of facts.\notes{said already}

\myheading{Compilation times and program sizes}
%
%
%
\begin{table}[t] 
  \small
  \centering
\begin{tabular}{@{~}l|r||r|r|r|r}
  Benchmark  & Original~ & Alda  & Compilation & Generated~ & Generated\\
  \Hex{3}name 
             & XSB size\,& size\,& time (ms)~\,& Python size & XSB size\,\\
  \noticlp{\hline}\iclp{\cline{1-6}}
  \iclp{\noalign{\vskip .4ex}}
  \noticlp{\hline}\iclp{\cline{1-6}}
  Join1      &  225 &  23 &  33.0 &  32 &   5\\\noticlp{\hline}\iclp{\cline{1-6}}
  Join2      &   41 &  11 &  18.5 &  16 &   9\\\noticlp{\hline}\iclp{\cline{1-6}}
  JoinDup    &  163 &  42 &  45.6 &  20 &  36\\\noticlp{\hline}\iclp{\cline{1-6}}  
  LUBM       &  377 & 125 & 116.4 &  29 & 110\\\noticlp{\hline}\iclp{\cline{1-6}}
  Mondial    &   36 &   8 &  16.2 &  16 &   6\\\noticlp{\hline}\iclp{\cline{1-6}}
  DBLP       &   20 &   4 &  16.3 &  16 &   2\\\noticlp{\hline}\iclp{\cline{1-6}} 
  TC         &   75 &   5 &   7.9 &  16 &   3\\\noticlp{\hline}\iclp{\cline{1-6}}
  TCrev      &  *75 &   5 &   7.7 &  16 &   3\\\noticlp{\hline}\iclp{\cline{1-6}}
  TCda       &   -- &   5 &   4.9 &  18 &   -\\\noticlp{\hline}\iclp{\cline{1-6}}
  TCpy       &   -- &   7 &   6.5 &  11 &   -\\\noticlp{\hline}\iclp{\cline{1-6}} 
  SG         &   90 &  13 &  17.9 &  20 &   7\\\noticlp{\hline}\iclp{\cline{1-6}}
  WordNet    &  298 &  58 &  76.2 &  44 &  28\\\noticlp{\hline}\iclp{\cline{1-6}}  
  Wine       & 1103 & 970 & 605.6 &  16 & 968\\\noticlp{\hline}\iclp{\cline{1-6}}
  ModSG      &   38 &  14 &  14.8 &  16 &  12\\\noticlp{\hline}\iclp{\cline{1-6}} 
  Win        &   24 &   4 &   9.5 &  16 &   2\\\noticlp{\hline}\iclp{\cline{1-6}}
  MagicSet   &   34 &   9 &  18.2 &  16 &   7\\\noticlp{\hline}\iclp{\cline{1-6}}
  ORBtimer   &   -- &  45 &  35.2 &  56 &   -\\\noticlp{\hline}\iclp{\cline{1-6}} 
  \iclp{\noalign{\vskip .4ex}}
  \noticlp{\hline}\iclp{\cline{1-6}}
  RBACnonloc &   -- & 423 & 346.4 & 538 &   4\\\noticlp{\hline}\iclp{\cline{1-6}}
  RBACallloc &   -- & 387 & 318.4 & 481 &   4\\\noticlp{\hline}\iclp{\cline{1-6}}
  RBACunion  &   -- & 386 & 316.6 & 481 &   3\\\noticlp{\hline}\iclp{\cline{1-6}}
  RBACda     &   -- & 385 & 312.8 & 483 &   -\\\noticlp{\hline}\iclp{\cline{1-6}}
  RBACpy     &   -- & 387 & 314.6 & 476 &   -\\\noticlp{\hline}\iclp{\cline{1-6}}
  RBACtimer  &   -- &  44 &  43.3 &  67 &   -\\\noticlp{\hline}\iclp{\cline{1-6}} 
  \iclp{\noalign{\vskip .4ex}}
  \noticlp{\hline}\iclp{\cline{1-6}}
  PA         &  *55 &  33 &  49.7 &  93 &   6\\\noticlp{\hline}\iclp{\cline{1-6}} 
  PAopt      &  *55 &  33 &  40.8 &  93 &   6\\\noticlp{\hline}\iclp{\cline{1-6}} 
  PAtimer    &   -- &  40 &  32.6 &  56 &   -\\\noticlp{\hline}\iclp{\cline{1-6}} 
  \iclp{\noalign{\vskip .4ex}}
  \noticlp{\hline}\iclp{\cline{1-6}}
\end{tabular}\Vex{1}
\def\tabcompilecaption{Compilation times and program sizes before and
  after compilation.}
\begin{tabular}{p{98ex}}
    For Original XSB size, entries without * are from OpenRuleBench,
    as in Table~\ref{tab-rulebench};
    * indicates XSB programs we added;
    -- means there is no corresponding XSB program.
    %
    For Generated XSB size, - means no XSB code is generated.\Vex{-1.75}
\end{tabular}
\caption{\tabcompilecaption}
\label{tab-compile}
\end{table}
\iclp{Table~\ref{tab-compile} shows Alda compilation times
  and related XSB, Alda, DistAlgo, and Python program sizes
before and after compilation. They are for all benchmarks described in
Section~\ref{sec-bench}, plus three variants of TC,
explained below in the paragraph on performance of classical queries 
using rules.
The Alda programs are 4--970 lines for OpenRuleBench benchmarks,
385--423 lines for RBAC benchmarks, and 33 lines for PA benchmarks.
For each group of benchmarks, there is a single shared Alda file of benchmarking code, shown in the last row of each group.

The compilation times for all programs are 0.6 seconds or less, and for all
but RBAC benchmarks and Wine in OpenRuleBench are about 0.1 seconds or less.
\noticlp{The generated Python files are 16--538 lines: the largest ones for RBAC
benchmarks, 93 lines for PA benchmarks, and all fewer than 50 lines for
benchmarks from OpenRuleBench.
The generated XSB files are all smaller than the Alda files, because they
contain only the rules, from just a few lines for many benchmarks to 968
lines for Wine from OpenRuleBench.}

For Alda programs that have corresponding XSB programs
(OpenRuleBench in Table~\ref{tab-rulebench} and PA), Alda programs are all
much smaller, almost all by dozens or even hundreds of lines, and by an
order of magnitude for Join1 and TC in OpenRuleBench, because we have all
the benchmarking code in the shared benchmarking file.
} 
\noticlp{
Table~\ref{tab-compile} shows the compilation times and program sizes
before and after compilation, for all three sets of benchmarks described in
Section~\ref{sec-bench}, plus three variants of TC in the first set as
explained in the subsequent Section~\ref{sec-expe-tc}.
For each group of benchmarks, there is a single shared Alda file of benchmarking code,
shown in the last row of each group;
for OpenRuleBench, ORBtimer includes 17 lines for pickling and timing of
pickling.

The compilation times for all benchmark programs are 0.6 seconds or less,
and for all but Wine and RBAC benchmarks about 0.1 seconds or less.

For OpenRuleBench benchmarks, Alda program sizes (4--970 lines) are all much smaller than
the corresponding XSB sizes (20--1103 lines), almost all by dozens or even hundreds of lines,
and by an order of magnitude for Join1 and TC.  In place of the extra XSB
code for benchmarking and manually added optimization directives, all Alda
programs use the single shared 45-line file, ORBtimer, for benchmarking
and pickling.
The generated XSB is even smaller (2--968 lines) and is exactly the number of rules
plus one line for \co{:- auto\_table.}, for each rule set.
The generated Python 
is all fewer than 50 lines, and is smaller for
benchmarks with fewer queries. 

For RBAC benchmarks, all Alda sizes (387--423 lines) include 3 files of 373 lines total for
all 9 RBAC classes; these 3 files take a total compilation time of 299.503
milliseconds,
generating a total compiled Python size of 456 lines.
Each way of computing the query functions is in a separate class extending
Hierarchical RBAC; RBACnonloc is over 30 lines more than others because all
query functions in Hierarchical RBAC, not just \co{authorizedUsers(role)},
are overridden to use field \co{transRH} in place of calls to
\co{transRH()}.
The generated Python is 476--538 lines, and the generated XSB programs contain only the 2--3 rules used and the \co{auto\_table} directive.

For PA benchmarks, 33 lines in Alda and 55 lines in XSB for each of PA and PAopt, the benchmarking code for the XSB programs is written in
a similar way as the benchmarks in OpenRuleBench, taking 23 lines each.
For each of PA and PAopt, the generated Python is 93 lines, and the generated XSB contains two programs, each containing the two rules used and the \co{auto\_table} directive.
} 

\myheading{Performance of classical queries using rules}
\label{sec-expe-tc}
\iclp{We experimented with four programs for computing transitive closure: 
TC, the TC benchmark from OpenRuleBench, 
which is the same as \co{trans\_rs} except with renamed predicates;
TCrev, a well-known variant with the two predicates reversed in the
recursive rule; 
and TCpy and TCda, which use the Python and DistAlgo \co{\WHILE} loops, 
respectively, in Section~\ref{sec-lang}.
For comparison, we also directly run the XSB program for TC from
OpenRuleBench, and its corresponding version for TCrev, except we change
\co{load\_dyn} to \co{load\_dync}, for much faster
reading of facts in XSB's canonical form; we call these programs TCXSB
and TCrevXSB, respectively.
The same data generation as OpenRuleBench is used to return large query results---almost
complete graphs.
} 
\noticlp{To evaluate the efficiency of classical queries using rules in Alda, we use
four programs for computing transitive closure: (1) TC---the TC benchmark
from OpenRuleBench, which is the same as \co{trans\_rs} except with renamed
predicates, (2) TCrev---same as \co{trans\_rs} but reversing the two
predicate names
in the recursive rule, which is a well-known variant,
(3) TCda---\co{\WHILE} loops with high-level queries
in Alda as in Section~\ref{sec-pred}, and (4) TCpy---\co{\WHILE} loops
with comprehensions in Python as in Section~\ref{sec-prob}.  

For comparison, we also directly run the XSB program for TC from
OpenRuleBench, and its corresponding version for TCrev, except we change
\co{load\_dyn} used in OpenRuleBench to \co{load\_dync}, for much faster
reading of facts in XSB's canonical form; we call these two programs TCXSB
and TCrevXSB, respectively.  Note that XSB programs in OpenRuleBench, not
using \co{load\_dync}, are significantly slower for large input, e.g., see
the DBLP benchmark in Section~\ref{sec-expe-scale}.


We use the data generator in OpenRuleBench to generate data, in the same way they use, which returns large query results.
%
The generator is sophisticated in trying to ensure randomness 
as well as cyclic vs.\ acyclic cases.
We use the same number of vertices, 1000, and a range of numbers of edges, 
10K to 100K, 
to include the first of two data points (50K and 500K edges)
reported in~\cite{Lia+09open}. 
For the cyclic graphs generated, even for the smallest data of 10K edges,
i.e., each vertex having edges going to only 1\% of vertices---10 out of
1000---on average, the resulting transitive closure is already the complete
graph of 1M edges.

Because of interfacing with XSB through files, the total running time of
Alda programs includes not only (1) reading data, (2) executing queries,
and (3) returning results, but also (2pre) preparing data, queries, and
commands and writing data
to files for XSB to start and read before (2),
and (2post) reading results from files written by XSB after (2).
We report the total running time as well as separate times for pickling
and for interfacing with XSB.

Figure~\ref{fig-TC} shows the running times of the TC benchmarks. %
RawR and PickleW are times for reading facts in XSB/Prolog form as used
in OpenRuleBench and writing them in pickled form for use in Alda,
respectively. Pickling is done only once; the pickled data is read in all
repeated runs and all of TC, TCrev, TCda, and TCpy.
%
TC\_extra and TCrev\_extra are the part of TC and TCrev, respectively, for
extra work interfacing with XSB, i.e., for 2pre and 2post and for XSB to
read data (xsbRdata) and write results (xsbWres).
Figure~\ref{fig-TC-break} shows the breakdown of TC\_extra and TCrev\_extra
among 2pre, 2post, xsbRdata, and xsbWres.
} 

%

%


\begin{figure}[t]
  \small
  \centering
\newcommand\figTC[1]{
\begin{tikzpicture}[every mark/.append style={mark size=2pt}]
  \begin{axis}[
    ymin=0, ymax=\noticlp{32}\iclp{50},
    xtick={10,20,30,40,50,60,70,80,90,100},
    xlabel={Number of edges (in thousands)},
    ylabel={CPU time (in seconds)},
    label style={font=\small},
    ylabel shift=-1ex,
    tick label style={font=\scriptsize},
    legend pos=north west,
    legend cell align={left},
    legend style={font=\scriptsize, row sep=-.5ex},
    ymajorgrids=true, 
    width=\arxiv{240pt}\oops{0.545\linewidth}\iclp{0.528\linewidth} 
    ]
\addplot[mark=pentagon,color=red] table[x=edges,y=TC,col sep=comma] {#1};
\addplot[mark=star,color=ForestGreen,densely dashed] table[x=edges,y=TCrev,col sep=comma] {#1};
\addplot[mark=x,color=ForestGreen,densely dashed] table[x=edges,y=TCrevXSB,col sep=comma] {#1};
\addplot[mark=diamond,color=red] table[x=edges,y=TCXSB,col sep=comma] {#1};
\addplot[mark=square,color=red] table[x=edges,y=TC_extra,col sep=comma] {#1};
\addplot[mark=+,color=ForestGreen,densely dashed] table[x=edges,y=TCrev_extra,col sep=comma] {#1};
\addplot[mark=triangle] table[x=edges,y=rawR,col sep=comma] {#1};
\addplot[mark=o] table[x=edges,y=pickleW,col sep=comma] {#1};
\legend{TC, TCrev, TCrevXSB, TCXSB, TC\_extra, TCrev\_extra\!, rawR, pickleW}
\end{axis}
\end{tikzpicture}
}
\figTC{TCcyc.csv}
\figTC{TCacyc.csv}\Vex{0}
\def\figtccaption{Running times of TC benchmarks on cyclic (left) and acyclic
  (right) graphs.}
\begin{tabular}{@{\!\!}p{102ex}}
  TCpy and TCda are not in the charts because they are
  asymptotically slower and took drastically longer: on 100 vertices, 
  with cyclic data of 200 edges (2\% of smallest data point in the charts),
  TCpy took 624.6 seconds, and TCda took 249.9 seconds; and with
  acyclic data of 600 edges, %
  TCpy took 160.4 seconds, and TCda took 65.2 seconds.\Vex{-1}
\end{tabular}
\caption{\figtccaption}
\label{fig-TC}
\Vex{1}
\end{figure}

\noticlp{
\pgfplotstableread{
edges 2pre 2post xsbRdata xsbWres rev2pre rev2post revxsbRdata revxsbWres
 10  0.078 0.737  0.060 0.780   0.082 0.718  0.057 0.820
 20  0.150 0.938  0.109 1.104   0.159 0.918  0.107 1.034
 30  0.207 1.012  0.158 1.199   0.172 1.043  0.168 1.259
 40  0.293 1.109  0.226 1.220   0.268 1.038  0.226 1.271
 50  0.302 1.106  0.280 1.258   0.354 1.100  0.280 1.413
 60  0.320 1.071  0.339 1.488   0.367 1.112  0.340 1.161
 70  0.466 1.145  0.419 1.366   0.473 1.124  0.424 1.315
 80  0.419 1.133  0.498 1.493   0.524 1.083  0.461 1.379
 90  0.532 1.112  0.535 1.358   0.566 1.103  0.496 1.348
100  0.592 1.135  0.640 1.382   0.570 1.139  0.554 1.435
}\TCexAcyc
\pgfplotstableread{
edges 2pre 2post xsbRdata xsbWres rev2pre rev2post revxsbRdata revxsbWres
 10  0.101 1.951  0.067 2.852   0.076 1.978  0.057 2.832
 20  0.138 2.018  0.113 2.706   0.136 1.938  0.114 2.768
 30  0.222 2.013  0.165 2.835   0.188 1.980  0.168 2.902
 40  0.221 1.999  0.215 2.792   0.275 1.962  0.218 2.865
 50  0.308 1.967  0.277 2.741   0.286 1.969  0.278 2.775
 60  0.380 1.964  0.343 2.851   0.384 1.984  0.341 2.851
 70  0.416 1.974  0.404 2.825   0.386 2.064  0.413 2.765
 80  0.493 2.015  0.480 2.879   0.518 1.975  0.474 2.755
 90  0.550 1.995  0.469 2.861   0.571 2.005  0.496 2.694
100  0.596 1.969  0.573 2.794   0.578 2.008  0.569 2.518
}\TCexCyc
\begin{figure}[ht]
  \centering
\newcommand\figTCex[1]{
\begin{tikzpicture}[
  /pgfplots/every axis/.style={
    ybar stacked, bar width=1ex,
    ymin=0, ymax=\noticlp{7.9}\iclp{9},
    xtick=data, xticklabels from table={#1}{edges},
    xlabel={Number of edges (in thousands)},
    ylabel={CPU time (in seconds)},
    label style={font=\small},
    ylabel shift=-1ex,
    tick label style={font=\scriptsize},
    legend style={cells={anchor=west}, legend pos=north west},
    legend style={font=\scriptsize, row sep=-.5ex},
    reverse legend=true,
    width=\arxiv{240pt}\oops{0.545\linewidth}\ickp{0.542\linewidth} 
  }
  ]
\begin{axis}[bar shift=-.25ex,
  ymajorgrids=true, 
  ]
  \addplot[fill=orange!60] table[y=2pre,meta=edges,x expr=\coordindex] {#1};
  \addplot[fill=green!60] table[y=xsbRdata,meta=edges,x expr=\coordindex]{#1};
  \addplot[fill=red!60] table[y=xsbWres,meta=edges,x expr=\coordindex] {#1};
  \addplot[fill=blue!60] table[y=2post,meta=edges,x expr=\coordindex] {#1};

  \legend{2pre, xsbRdata\!, xsbWres, 2post}
  \node[draw,fill=white,anchor=north east] at (rel axis cs: 0.97,0.97){\shortstack[l]{
      {\scriptsize {\bf Bar pairs:}}\\
      {\scriptsize left: TC\_extra}\\
      {\scriptsize right: TCrev\_extra}
    }};
\end{axis}
\begin{axis}[bar shift=.75ex]
  \addplot[fill=orange!60] table[y=rev2pre,meta=edges,x expr=\coordindex] {#1};
  \addplot[fill=green!60] table[y=revxsbRdata,meta=edges,x expr=\coordindex]{#1};
  \addplot[fill=red!60] table[y=revxsbWres,meta=edges,x expr=\coordindex] {#1};
  \addplot[fill=blue!60] table[y=rev2post,meta=edges,x expr=\coordindex] {#1};
\end{axis}
\end{tikzpicture}\Vex{-1}
}
\figTCex{\TCexCyc}
\figTCex{\TCexAcyc}
\caption{Breakdown of TC\_extra and TCrev\_extra.}
\label{fig-TC-break}
\end{figure}
} 

\noticlp{
Not shown in Figure~\ref{fig-TC}, but the times in TC and TCrev for reading
facts are similar to PickleW; 
the times in TCXSB and TCrevXSB for reading facts are similar to RawR.
The remaining times in TCXSB and TCrevXSB are XSB query times
only, because XSB programs in OpenRuleBench do not output or even collect
the query result in any way.

Also not shown in Figures~\ref{fig-TC} and~\ref{fig-TC-break} are the times
in TC and TCrev for starting the XSB process and, as part of 2pre, for
preparing the queries to start XSB with proper arguments and status checks.
These are small, at 0.1--0.2 seconds and 0.03--0.04 seconds, respectively,
because XSB is invoked only once in each run.

We observe the following results, which are all as expected:
\begin{itemize}
\item The extra times interfacing with XSB are obvious (up to 5.9 seconds out of 29.2 total, for graphs of 100K edges), 
here dominated by passing query results by files (xsbWres and 2post),
as shown in Figure~\ref{fig-TC-break},
because the results of transitive closure are generally large,
and larger for cyclic graphs.
This overhead of going through files is expected
to be reduced to 1\% of it when using direct
mapping between XSB and Python data structures in memory.

\item The remaining times without the interfacing overhead are basically
  all XSB query times for the different XSB programs.
In particular, 
TCrev is faster than TC in Alda, 
but TCXSB is faster than TCrevXSB.
This is because OpenRuleBench %
uses the fastest manually optimized program for each problem,
which is TCXSB with subsumptive tabling for this specific benchmark,
while Alda-generated XSB programs use \co{auto\_table}, which is
variant tabling.
These 
are known to cause the observed performance differences~\cite{TekLiu10RuleQuery-PPDP,TekLiu11RuleQueryBeat-SIGMOD}.
Alda compiler can be extended to automatically generate optimal
programs using previously studied
methods~\cite{LiuSto09Rules-TOPLAS,Tek+08RulePE-AMAST,TekLiu10RuleQuery-PPDP,TekLiu11RuleQueryBeat-SIGMOD}.

\item TCpy and TCda, while being drastically easier to write than low-level
  code despite not as easy as rules, are exceedingly inefficient. In
  contrast, TC and TCrev that use rules are drastically faster.
\end{itemize}
Note that both TC and TCrev in Alda, even including the extra times
interfacing with XSB, are faster or even drastically faster than all
systems reported in OpenRuleBench~\cite{Lia+09open} except for XSB and one or
two other systems (for 50K edges, 17.0 seconds for TC in Alda vs.\ up to 184.0
seconds and even an error on cyclic data, and 6.8 seconds for TC in Alda vs.\
up to 120.4 seconds on acyclic data, where the XSB query times were similar
as in our measurements; note that these are despite OpenRuleBench reporting
only the times for queries, not reading data or writing results).
This is also despite each of those programs having been manually optimized in the
most advantageous and beneficial ways for each system~\cite{Lia+09open}.
} 
\iclp{
Figure~\ref{fig-TC} shows the running times of the TC benchmarks. %
RawR and PickleW are times for reading facts in XSB/Prolog form as used
in OpenRuleBench and writing them in pickled form for use in Alda,
respectively. Pickling is done only once; the pickled data is read in all
repeated runs and all of TC, TCrev, TCda, and TCpy.
The running time of Alda programs includes not only (1) reading data, 
(2) executing queries, and (3) returning results, but also 
(2pre) preparing data, queries, and commands and writing data to files for XSB 
before (2), and (2post) reading results from files written by XSB after (2).
TC\_extra and TCrev\_extra are the part of TC and TCrev, respectively, for
extra work interfacing with XSB, i.e., for 2pre and 2post and for XSB to read data 
(xsbRdata) and write results (xsbWres).
A breakdown showing the time needed for each step in the extra work is in~\cite{Liu+22RuleLang-arxiv}.

The results are as expected: the two Alda programs that use rules are
asymptotically, drastically faster than Python and DistAlgo \co{\WHILE} loops,
and they exhibit known notable performance
differences~\cite{TekLiu10RuleQuery-PPDP,TekLiu11RuleQueryBeat-SIGMOD}.
Most notably but as expected, passing the query results back from XSB has a
high overhead, up to 5.9 seconds, out of 29.2 seconds total, for graphs of 100K
edges, but this overhead is expected to be reduced to 1\% of it when the
in-memory Python-XSB interface is used.
Note that reversing the two predicates in the recursive rule
does give a linear-factor asymptotically different running time,
but that barely shows because OpenRuleBench uses almost complete graphs.
}

\myheading{Integrating with objects, updates, and set queries}
\label{sec-expe-rbac}
\iclp{We use RBAC benchmarks in Table~\ref{tab-rbac}, Section~\ref{sec-rbac}, for
this evaluation, especially with frequent queries and updates and
intensively frequent restart of XSB for queries randomly mixed with updates
of the queried data: 5000 users, 500 roles, 5000 UR relation, 550 RH
relation, up to 500 queries, and 230 updates of various kinds.

Figure~\ref{fig-RBAC} shows the running times of the RBAC benchmarks, all
scaling linearly in the number of queries, as
expected.
Labels with suffix \_extra indicate the part of the running time of the
corresponding program for extra work interfacing with XSB.
A breakdown of the time for extra work is in \cite{Liu+22RuleLang-arxiv}.

The results are as expected as well: RBACunion and RBACalloc are very close,
and are much slower than RBACnonloc---up to 331.7 and 333.9 seconds, respectively, vs.\
97.9 seconds.
Most notably but as expected, the overhead of repeated queries using XSB is
high for RBACunion and RBACalloc, but low for RBACnonloc, up to 134.5 and 145.9
seconds, respectively, vs.\ 2.8 seconds. The highest overhead is from restarting XSB 500 times,
which will be totally eliminated when in-memory interface is used.
} 
\noticlp{To evaluate the performance of using rules with objects and updates, and
of different ways of using rules as well as not using rules, we use the
RBAC benchmarks in Section~\ref{sec-rbac}.
We especially consider cases with frequent queries and updates and intensively
frequent restart of XSB for queries randomly mixed with updates of the queried data. 
%
%
%
%
%
%
%

%
%
We create 5000 users and 500 roles, 
and randomly generate a user-role assignment \co{UR} of size
5500 
with a maximum of 10 roles per user, 
and a role hierarchy \co{RH} of size
550 
and height 5.  
We run the following set of workloads: iterate and randomly do one of the
following operations in each iteration: add/delete user (50 total each of
add and delete), add/delete role (5 total each), add/delete \co{UR} pair
(55 total each), add/delete \co{RH} pair (5 total each), and query
authorized users ($n$ total), for $n$ up to 500 at intervals of 50.  We
measure the running time of the workload for each $n$.
%
} 
\oops{
} 


\pgfplotstableread{
query RBACunion RBACallloc RBACnonloc RBACunionEx RBACalllocEx RBACnonlocEx
 50   33.325   33.448  14.579          13.476     14.575  2.519
100   66.529   66.534  23.865          26.880	  29.022  2.430
150   99.523  100.032  33.591          40.054	  43.608  2.760
200  132.518  133.369  42.837          53.327	  58.314  2.745
250  165.922  166.691  51.969          66.903	  72.703  2.616
300  198.753  199.152  61.801          80.052	  86.409  2.777
350  232.440  233.429  70.418          93.513	 102.006  2.752
400  265.139  266.651  79.941         106.860	 116.418  2.813
450  298.280  298.790  89.045         120.200	 130.043  2.846
500  331.697  333.851  97.923         134.497	 145.940  2.831
}\RBACdata
\begin{figure}[t!]
  \centering
\begin{tikzpicture}[every mark/.append style={mark size=2pt}]
  \begin{axis}[
    ymin=0, ymax=\noticlp{384}\iclp{490},
    xtick=data, xticklabels from table={\RBACdata}{query},
    xlabel={Number of \co{AuthorizedUsers} queries},
    ylabel={CPU time (in seconds)},
    label style={font=\small},
    ylabel shift=-1ex,
    tick label style={font=\scriptsize},
    mark options={solid},
    legend style={cells={anchor=west}, legend pos=north west},
    legend style={font=\scriptsize, row sep=-.5ex},
    ymajorgrids=true,
    width=\arxiv{240pt}\iclp{0.545\linewidth} 
]
\addplot[mark=square,color=red] table[x=query,y=RBACallloc] {\RBACdata};
\addplot[mark=|,color=blue,densely dashed] table[x=query,y=RBACunion] {\RBACdata};
\addplot[mark=diamond,color=red] table[x=query,y=RBACalllocEx] {\RBACdata};
\addplot[mark=star,color=blue,densely dashed] table[x=query,y=RBACunionEx] {\RBACdata};
\addplot[mark=o,color=ForestGreen,dash pattern=on 4.5pt off 2pt] table[x=query,y=RBACnonloc] {\RBACdata};
\addplot[mark=triangle,color=ForestGreen,dash pattern=on 4.5pt off 2pt] table[x=query,y=RBACnonlocEx] {\RBACdata};

\legend{RBACallloc, RBACunion, RBACallloc\_extra, RBACunion\_extra, 
  RBACnonloc, RBACnonloc\_extra}
\end{axis}
\end{tikzpicture}\Vex{-1}
\Hex{5}
{\footnotesize
\begin{tabular}[b]{c}
\fbox{
\begin{tabular}{rr}
RBACpy & \\
\#queries & time (s)\\
 50 &  688.7\\
100 & 1381.5 \\
150 & > 30 min\\
\\
RBACda & \\
\#queries & time (s)\\
 50 &  384.6\\
100 &  768.1\\
150 & 1141.7\\
200 & 1517.5\\
250 & > 30 min\\
\end{tabular}}\Vex{1}\\
\begin{tabular}{@{~}p{29ex}}
RBACpy and RBACda are
not in the chart because
they are drastically slower.
\end{tabular}
\Vex{1}
\end{tabular}\Hex{4}
}
\caption{Running times of RBAC benchmarks, for a workload of updates and
  queries over 5000 users, 500 roles, 5500 user-role assignments, and 550
  role hierarchy pairs.  
  \noticlp{
  RBACpy and RBACda are not in the chart
  because they are much slower: on data point 50,
  they are 
  688.7 and 384.6 seconds,
  respectively, but they increased linearly 
  as expected,
  to 
  1381.5 seconds on data point 100 and to 1517.5 seconds on data
  point 200, respectively, and failed to complete by the time limit of
  30 minutes on larger data points.\Vex{-1}}}
\label{fig-RBAC}
\end{figure}

\noticlp{
\pgfplotstableread{
query   2pre    Wdata  prepStart  xsbStart xsbRdata xsbWres  2post
 50	 2.935	0.181   2.754	   5.184   0.288     2.708    2.315
100	 5.854	0.386   5.468	  10.417   0.546     5.442    4.595
150	 8.756	0.591   8.165	  15.488   0.814     8.220    6.830
200	11.591	0.750  10.841	  20.725   1.047    11.013    9.123
250	14.672	0.971  13.701	  25.793   1.329    13.736   11.409
300	17.458	1.146  16.312	  31.033   1.604    16.277   13.645
350	20.432	1.387  19.045	  36.237   1.888    19.190   15.911
400	23.338	1.524  21.814	  41.204   2.119    21.865   18.385
450	26.304	1.765  24.539	  46.382   2.389    24.530   20.586
500	29.393	1.915  27.478	  51.987   2.657    27.466   22.982
}\RBACexUnion
\pgfplotstableread{
query   2pre    Wdata  prepStart  xsbStart xsbRdata xsbWres  2post
 50	 2.989	0.281   2.708	   5.224   0.369     2.708    3.285
100	 5.964	0.522   5.442	  10.381   0.729     5.442    6.506
150	 8.986	0.766   8.220	  15.537   1.104     8.220    9.761
200	12.088	1.075  11.013	  20.727   1.454    11.013   13.031
250	15.067	1.331  13.736	  25.878   1.766    13.736   16.256
300	17.825	1.548  16.277	  30.801   2.135    16.277   19.371
350	21.011	1.821  19.190	  36.471   2.526    19.190   22.808
400	24.019	2.154  21.865	  41.387   2.893    21.865   26.256
450	26.814	2.284  24.530	  46.298   3.193    24.530   29.208
500	30.182	2.716  27.466	  51.797   3.590    27.466   32.905
}\RBACexAllloc
\pgfplotstableread{
query   2pre   Wdata  prepStart	 xsbStart xsbRdata xsbWres  2post
 50	0.492  0.033  0.459	 0.963	  0.077	   0.459    0.528
100	0.471  0.034  0.437	 0.926	  0.087	   0.437    0.508
150	0.549  0.042  0.507	 1.084	  0.085	   0.507    0.535
200	0.542  0.036  0.506	 1.058	  0.093	   0.506    0.546
250	0.502  0.033  0.469	 1.027	  0.082	   0.469    0.536
300	0.543  0.037  0.506	 1.065	  0.095	   0.506    0.568
350	0.545  0.04   0.505	 1.079	  0.088	   0.505    0.535
400	0.549  0.039  0.51	 1.077	  0.089	   0.510    0.587
450	0.564  0.048  0.516	 1.101	  0.088	   0.516    0.577
500	0.552  0.035  0.517	 1.094	  0.100	   0.517    0.568
}\RBACexNonloc
\begin{figure}[t]
  \centering
\begin{tikzpicture}[
  /pgfplots/every axis/.style={
    ybar stacked, bar width=1ex,
    ymin=0, ymax=\noticlp{189}\iclp{220},
    xlabel={Number of \co{AuthorizedUsers} queries},
    ylabel={CPU time (in seconds)},
    label style={font=\small},
    ylabel shift=-1ex,
    tick label style={font=\scriptsize},
    legend style={cells={anchor=west}, legend pos=north west},
    legend style={font=\scriptsize, row sep=-.5ex},
    reverse legend=true,
    ymajorgrids=true, 
    height=\arxiv{40}\oops{45}\iclp{40}ex, 
    width=\noticlp{0.55}\iclp{.66}\linewidth
  }
  ]
\newcommand{\axisRBACex}[3]{
\begin{axis}[bar shift=#2]
  \addplot[fill=yellow!#3] table[y=prepStart,meta=query,x expr=\coordindex] {#1};
  \addplot[fill=orange!#3] table[y=Wdata,meta=query,x expr=\coordindex] {#1};

  \addplot[fill=pink!#3] table[y=xsbStart,meta=query,x expr=\coordindex]{#1};
  \addplot[fill=green!#3] table[y=xsbRdata,meta=query,x expr=\coordindex]{#1};
  \addplot[fill=red!#3] table[y=xsbWres,meta=query,x expr=\coordindex] {#1};
  \addplot[fill=blue!#3] table[y=2post,meta=query,x expr=\coordindex] {#1};
  \legend{2pre\_prepStart\!\!,2pre\_rest, xsbStart, xsbRdata\!,xsbWres, 2post}

  \node[draw,fill=white,anchor=north east] at (rel axis cs: 0.97,0.97){\shortstack[l]{
      {\scriptsize \Hex{-1}{\bf Bar triples:}}\\
      {\scriptsize \Hex{-1}left: RBACallloc\_extra}\\
      {\scriptsize \Hex{-1}middle: RBACunion\_extra}\\
      {\scriptsize \Hex{-1}right: RBACnonloc\_extra}
    }};
\end{axis}
}
\axisRBACex{\RBACexNonloc}{1ex}{60}
\axisRBACex{\RBACexUnion}{0ex}{60}
\axisRBACex{\RBACexAllloc}{-1ex}{60}
\end{tikzpicture}\Vex{-1}
\caption{Breakdown of RBACallloc\_extra, RBACunion\_extra, and
  RBACnonloc\_extra.}
\label{fig-RBAC-break}
\end{figure}
} 

\noticlp{Figure~\ref{fig-RBAC} shows the running times of the RBAC benchmarks, all
scaling linearly in the number of \co{AuthorizedUsers} queries, as
expected.
Labels with suffix \_extra indicate the part of the running time of the
corresponding program for extra work interfacing with XSB: those as in Figure~\ref{fig-TC-break} plus here the times
for starting XSB processes.
Figure~\ref{fig-RBAC-break} shows the breakdown of the times for the extra
work, separating also the part of 2pre on preparing the queries and
commands to start XSB
(2pre\_prepStart) from the rest of 2pre on writing data to
files for XSB to read (2pre\_rest) and from the time for starting the XSB process (xsbStart).

We observe the following results, which are all as expected as well:
\begin{itemize}

\item The extra times interfacing with XSB are again obvious (up to 145.9 and 134.5 seconds for RBACallloc and RBACunion, respectively, though only 2.8 seconds for RBACnonloc), 
  here dominated mostly by preparing XSB queries and command lines and starting
  XSB, as shown in Figure~\ref{fig-RBAC-break}. This is unlike for TC benchmarks,
  because the data and results are much smaller but all the work associated
  with invoking XSB through command line is repeated (up to 500 times for 500 \co{AuthorizedUsers} queries for RBACalloc and RBACunion, 
  and 20 times for 20 updates to \co{ROLES} and \co{RH} for RBACnonloc).  
  This overhead from repeatedly preparing and restarting XSB will be totally eliminated
  when the in-memory interface between XSB and Python is used.
  
\item RBACallloc and RBACunion are very close (333.9 and 331.7 seconds, respectively), as shown in
  Figure~\ref{fig-RBAC}, with a slightly higher interfacing overhead by
  RBACallloc as expected for the extra data and results passed due to
  \co{ROLES}, but compensated by slightly faster queries in XSB than set
  union operations in Python.
  %
  RBACnonloc (97.9 seconds) is the fastest, more than 3 times as fast as RBACallloc and RBACunion,
  because the inference for computing \co{transRH} is done at updates not
  queries, and there is a smaller, fixed number of updates.
  Its performance can be optimized even more with incremental computation,
  as for either set queries,
  e.g.,~\cite{Liu+06ImplCRBAC-PEPM,Gor+12Compose-PEPM,Liu+16IncOQ-PPDP}, or
  logic rules, e.g.,~\cite{SahaRam03}.

\item RBACpy and RBACda are again exceedingly inefficient, as expected.
  In contrast, the three programs that use rules are all significantly
  faster.
\end{itemize}

%


%
%
} 

\myheading{Integrating with aggregate queries and recursive functions}
\label{sec-expe-pa}
\iclp{We use PA and PAopt benchmarks and their corresponding programs in XSB, 
as described in Section~\ref{sec-pa}, for this evaluation, 
and we focus on applying the analysis to large programs as input data.  The
programs analyzed include 9 widely-used open-source Python packages for a
wide range of application domains: NumPy, SciPy, MatPlotLib, Pandas, SymPy, 
Blender, Django, Scikit-learn, and PyTorch---with 641K--5.1M input facts
total and 252K--2.2M facts used by the analysis.

Table~\ref{tab-PA-expe} shows data sizes,
analysis results, and running times of the analysis.
The columns are sorted by the total number of facts used.
A breakdown of the running time into steps interfacing with
XSB as well as the remainder of the running time is in \cite{Liu+22RuleLang-arxiv}.

The results are not as expected: we found the corresponding XSB
programs to be highly inefficient, being all slower and even drastically
slower than Alda programs, even 120 times slower for PyTorch.
Significant effort was spent on performance debugging and manual
optimization, and we eventually created a version that is faster than Alda---5.1
seconds vs.\ 15.2 seconds on the largest input, SymPy---by using additional
directives for targeted tabling that also subsumes some indexing.

As expected, the Alda programs here have a high overhead of 
passing the data to XSB, up to 13.1 seconds on SymPy, which again is expected to be reduced
to 1\% of it with in-memory Python-XSB interface. This means that the
resulting Alda programs would be faster than even the manually optimized XSB, showing that computations
not using rules, e.g., aggregations and functions, are not only simpler and
easier in Alda/Python than in XSB but also faster.
} 
\noticlp{To evaluate the performance of integrated use of rules with aggregate
queries and recursive functions, we use two benchmarks for class hierarchy
analysis: PA and PAopt, and the corresponding programs in XSB, as described
in Section~\ref{sec-pa},
and we focus on applying the analysis to large input programs.
 
We found the XSB programs corresponding to PA and PAopt, which we call
PAXSB and PAoptXSB, respectively, to be highly inefficient, being slower
and even drastically slower than Alda programs.
%
We tried many manual optimizations by manipulating the rules and adding
directives, including with help from an XSB expert, and eventually found 
a version that is faster than the Alda programs, which we call PAXSBopt, 
that uses additional directives for
targeted tabling that also subsumes some indexing.

The programs analyzed 
include 9 widely-used open-source Python packages, 
all available on GitHub (\myurl{https://github.com/}):
NumPy (\ver{v1.21.5}\ghurl{https://github.com/}{numpy/numpy}) and 
SciPy (\ver{v1.7.3}\ghurl{https://github.com/}{scipy/scipy}), for scientific computation;
MatPlotLib (matplot) (\ver{v3.5.1}\ghurl{https://github.com/}{matplotlib/matplotlib}), for visualization;
Pandas (\ver{v1.3.5}\ghurl{https://github.com/}{pandas-dev/pandas}), for data analysis; 
SymPy (\ver{sympy-1.9}\ghurl{https://github.com/}{sympy/sympy}), for symbolic computation; 
Django (\ver{4.0}\ghurl{https://github.com/}{django/django}), for web development; 
Scikit-learn (sklearn) (\ver{1.0.1}\ghurl{https://github.com/}{scikit-learn/scikit-learn}) 
and
PyTorch (\ver{v1.10.1}\ghurl{https://github.com/}{pytorch/pytorch}), for machine learning;
and 
Blender (\ver{v3.0.0}\ghurl{https://github.com/}{blender/blender}), for 3D graphics.
%
Each of these Python packages 
contains many files and directories.
We first parse each file and translate the resulting abstract
syntax tree (AST) along with file and directory information into Datalog
facts.
We then run the benchmarks.
} 

\begin{table*}[t]
  \footnotesize
  \centering
\begin{tabular}{@{\,}l@{\,} @{\,}l@{\,} ||@{\,}r@{\,}|@{\,}r@{\,}|@{\,}r@{\,}|@{\,}r@{\,}|@{\,}r@{\,}|@{\,}r@{\,}|@{\,}r@{\,}|@{\,}r@{\,}|@{\,}r@{\,}}
Measure & Item/Name     & numpy   & django  & sklearn & blender & pandas  & matplot & scipy   & pytorch   & sympy \\
    \noticlp{\hline}\iclp{\cline{1-11}}\noticlp{\hline}\iclp{\cline{1-11}}
Data	& Total	        & 640,715 & 815,551 & 862,031 & 909,600 & 942,315 & 1,064,859 & 1,092,466 & 5,142,905 & 5,115,105 \\
size	& \co{ClassDef}	& 587	  & 1,835   & 535     & 2,146	& 849	  & 994	    & 898     & 6,467	  & 1,830 \\
	& \co{Name}	& 96,076  & 119,077 & 137,066 & 107,638	& 153,664 & 152,357 & 178,754 & 797,072	  & 1,063,842 \\
 	& \co{Member}	& 155,207 & 199,416 & 210,410 & 242,531	& 227,766 & 268,736 & 260,848 & 1,270,917 & 1,112,296 \\
 	& Total used    & 251,870 & 320,328 & 348,011 & 352,315 & 382,279 & 422,087 & 440,500 & 2,074,456 & 2,177,968 \\\noticlp{\hline}\iclp{\cline{1-11}}
Ratio   & Used/Total    & 39.3\%  & 39.3\%  & 40.4\%  & 38.7\%  & 40.6\%  & 39.6\%  & 40.3\%  & 40.3\%    & 42.6\% \\
    \noticlp{\hline}\iclp{\cline{1-11}}\noticlp{\hline}\iclp{\cline{1-11}}

Result  & \co{\#defined}       & 519 &  1610 &   533 &  2118 &   804 &   935 &   882 &  4323 &  1786 \\
size    & \co{\#extending}     & 419 &  1457 &   710 &  2951 &   407 &   610 &   719 &  2207 &  1816 \\
        & \co{\#roots}         &  79 &   225 &    51 &   133 &    88 &   104 &    60 &   137 &    92 \\
        & \co{max\_height}     &   8 &     7 &     5 &     4 &     7 &     5 &     3 &     5 &    12 \\
        & \co{\#roots\_max\_h} &   1 &     2 &     2 &     1 &     2 &     1 &     4 &     1 &     1 \\
        & \co{\#desc}          & 427 &  2329 &   822 &  4376 &   436 &   605 &   721 &  2174 &  2413 \\
        & \co{max\_desc}       &  84 &   309 &   256 & 1,638 &    65 &    47 &   353 & 1,045 & 1,078 \\
        & \co{\#roots\_max\_d} &   1 &     1 &     1 &     1 &     1 &     1 &     1 &     1 &     1 \\
    \noticlp{\hline}\iclp{\cline{1-11}}\noticlp{\hline}\iclp{\cline{1-11}}



Running & PA       &  2.542 &   3.573 &  3.263 &   5.134 &  3.342 &  3.733 &  3.646 &  14.652 &  15.243 \\
time    & PAopt    &  2.631 &   3.520 &  3.235 &   4.661 &  3.341 &  3.676 &  3.633 &  14.706 &  15.132 \\
(in     & PAXSB    &  6.297 & 112.091 & 10.795 & 243.765 &  6.378 & 14.400 & 22.221 & 969.228 &  65.382 \\
seconds)& PAoptXSB & 13.170 & 343.428 & 17.066 & 326.629 & 18.863 & 40.871 & 29.675 &1773.374 & 181.961 \\
        & PAXSBopt &  0.804 &   1.499 &  1.071 &   2.149 &  1.118 &  1.317 &  1.300 &   5.158 &   5.051 \\
    \noticlp{\hline}\iclp{\cline{1-11}}
Ratio	& PAopt    &103.5\% &  98.5\% & 99.1\% &  90.8\% &100.0\% &  98.5\% & 99.6\% &  100.4\% &  99.3\% \\
over	& PAXSB    &247.7\% &3137.2\% &330.8\% &4748.0\% &190.8\% & 385.7\% &609.5\% & 6615.0\% & 428.9\% \\ 
PA	& PAoptXSB &518.1\% &9611.7\% &523.0\% &6362.1\% &564.4\% &1094.9\% &813.9\% &12103.3\% &1193.7\% \\
	& PAXSBopt & 31.6\% &  42.0\% & 32.8\% &  41.9\% & 33.5\% &  35.3\% & 35.6\% &   35.2\% &  33.1\% \\
    \noticlp{\hline}\iclp{\cline{1-11}}\noticlp{\hline}\iclp{\cline{1-11}}
\end{tabular}\Vex{1}
\def\tabpaexpecaption{Data size, analysis results, and running times 
  for program analysis benchmarks.}
\begin{tabular}{@{}p{112ex}}
  Total is the total number of facts about each package.
  Total used is the sum of numbers of \co{ClassDef}, \co{Name},
  and \co{Member} facts.
\end{tabular}\Vex{-3}
\caption{\tabpaexpecaption}
\label{tab-PA-expe}
\end{table*}
\noticlp{Table~\ref{tab-PA-expe} shows data sizes,
analysis results, and running times of the analysis.
The columns are sorted by the total number of facts used (i.e., all
\co{ClassDef}, \co{Name}, and \co{Member} facts, 252K--2.2M)), which mostly coincides
with the total number of facts (641K--5.1M) except for the largest two, \co{pytorch} and
\co{sympy}.
Note that, even for the small rule set \co{class\_extends\_rs}, the
total number of facts used is already 38.7--42.6\% of the total number of
facts, because \co{Member} and \co{Name} are two of the largest.
Figure~\ref{fig-PA-break} shows the breakdown of the time interfacing with
XSB (2pre, 2post, and xsbRdata, as in Figure~\ref{fig-TC-break} except that
xsbWres is even smaller than 2post and is not shown) plus the remaining
time in the total time for PA and PAopt (total\_rest).
} 
\noticlp{
\pgfplotstableread{
lib  2pre  2post xsbRdata     total   rest
numpy    1.270 0.025 0.833     2.542  0.415
django   1.425 0.137 0.953     3.573  1.058
sklearn  1.553 0.051 1.147     3.263  0.512
blender  1.485 0.220 1.152     5.134  2.277
pandas   1.634 0.030 1.185     3.342  0.493
matplot  1.832 0.043 1.315     3.733  0.543
scipy    1.831 0.046 1.216     3.646  0.554
pytorch  7.386 0.204 4.684    14.652  2.378
sympy    7.908 0.152 5.030    15.243  2.153
}\PAex
\pgfplotstableread{
lib  2pre  2post xsbRdata     total   rest
numpy    1.289 0.027 0.917     2.631  0.398
django   1.454 0.14  0.988     3.52   0.938
sklearn  1.589 0.053 1.102     3.235  0.491
blender  1.595 0.213 1.146     4.661  1.707
pandas   1.650 0.032 1.170     3.341  0.489
matplot  1.791 0.042 1.318     3.676  0.525
scipy    1.837 0.045 1.218     3.633  0.533
pytorch  7.578 0.205 4.651    14.706  2.272
sympy    7.981 0.143 5.094    15.132  1.914
}\PAoptex
\begin{figure}[t]
  \centering
\begin{tikzpicture}[
  /pgfplots/every axis/.style={
    ybar stacked, bar width=1ex,
    ymin=0, ymax=16,
    xtick=data, xticklabels from table={\PAex}{lib},
    xlabel={Number of edges (in thousands)},
    ylabel={CPU time (in seconds)},
    label style={font=\small},
    ylabel shift=-1ex,
    tick label style={font=\scriptsize},
    x tick label style={at={(axis description cs:2.01,+0.0)},rotate=-30,anchor=west},
    legend style={cells={anchor=west}, legend pos=north west},
    legend style={font=\scriptsize, row sep=-.5ex},
    reverse legend=true,
    ymajorgrids=true, 
    width=\arxiv{240pt}\iclp{0.545\linewidth} 
  }
  ]
\newcommand{\axisPAex}[3]{
\begin{axis}[bar shift=#2]
  \addplot[fill=orange!#3] table[y=2pre,meta=lib,x expr=\coordindex] {#1};
  \addplot[fill=green!#3] table[y=xsbRdata,meta=lib,x expr=\coordindex]{#1};
  \addplot[fill=blue!#3] table[y=2post,meta=lib,x expr=\coordindex] {#1};
  \addplot[fill=white!#3] table[y=rest,meta=lib,x expr=\coordindex] {#1};
  \legend{2pre, xsbRdata\!, 2post, total\_rest}
  \node[draw,fill=white,anchor=north east] at (rel axis cs: 0.78,0.97){\shortstack[l]{
      {\scriptsize {\bf Bar pairs:}}\\
      {\scriptsize left: PA}\\
      {\scriptsize right: PAopt}
    }};
\end{axis}
}
\axisPAex{\PAoptex}{1ex}{60}
\axisPAex{\PAex}{0ex}{60}
\end{tikzpicture}\Vex{-1}
\caption{Running times of PA and PAopt.}
\label{fig-PA-break}
\end{figure}
} 
%
%

\noticlp{We observe the following results, where the performance of the corresponding XSB programs are entirely unexpected:
\begin{itemize}
\item The times interfacing with XSB is again obvious (up to 13.1 seconds out of 15.2 total for PA), here vastly
  dominated by the time to pass AST facts to XSB as shown in
  Figure~\ref{fig-PA-break}, because of the large data sizes vs.\ the small
  result sizes shown in Table~\ref{tab-PA-expe}.  This contrasts the times
  dominated by passing results in Figure~\ref{fig-TC-break} for TC benchmarks and by repeated
  starting of XSB in Figure~\ref{fig-RBAC-break} for RBAC benchmarks; note that XSB is invoked only twice
  here, once for each rule set in the benchmark.  Again, this
  overhead is expected to be reduced to 1\% of it with in-memory mapping between XSB and Python data
  structures.

\item The running times of PA and PAopt are similar and mostly increase as
  the data sizes increase, as shown in Table~\ref{tab-PA-expe} and
  Figure~\ref{fig-PA-break}.
  PAopt is in most cases (all but \co{numpy} and \co{pytorch}) very
  slightly faster than PA, because querying using rules takes only a small
  part of the total time (0.5--11.6\% of PA, and 0.3--2.5\% of PAopt), with
  the rest on interfacing with XSB and on other queries using functions and
  aggregations.
  Querying using rules in PAopt is actually 1.4--11.8 times as fast as that
  in PA.

\item PAXSB and PAoptXSB have vastly varying running times, as shown in
  Table~\ref{tab-PA-expe}, unlike PA and PAopt, and are much slower than PA
  and PAopt, taking 1.9--121.1 times as long as PA and even more than PAopt.
  This is after we already manually added tabling for \co{height} and
  \co{num\_desc} to match PA and PAopt, after finding that \co{auto\_table}
  only tabled predicate \co{desc}.
  Note that PAXSB and PAoptXSB are mostly slower for larger result
  sizes, as opposed to input sizes, though all result sizes are orders of
  magnitude smaller than input sizes.
  %

%
  PAXSBopt, with manual optimizations after trying various combinations of
  tabling, indexing, and rewriting for the remaining predicates, is
  58.0-78.4\%
  faster that PA (5.1 vs.\ 15.2 seconds on the largest input, SymPy).  Again, previously studied
  methods~\cite{TekLiu10RuleQuery-PPDP,TekLiu11RuleQueryBeat-SIGMOD} can be
  added to the Alda compiler to automatically generate optimal tabling and
  indexing directives as needed; manually applying these sophisticated
  methods is too tedious.

\end{itemize}
Note that these results also mean that, with the high overhead of interfacing XSB drastically reduced, the resulting Alda programs would be faster than even the manually optimized XSB, showing that computations not
using rules, e.g., aggregations and functions, are not only simpler and easier in Alda/Python than
in XSB but also faster.
} 

\myheading{Scaling with data and rules}
\label{sec-expe-scale}
\iclp{We use the two largest benchmarks from OpenRuleBench: DBLP, with over 2.4 million facts,
the largest real-world data set among all in OpenRuleBench; and Wine, with
961 rules, the largest rule set among all.
Table~\ref{tab-dblp-wine} shows the running times for both 
benchmarks, for both the Alda programs and the XSB programs.
\_extra is the part of the total time on 2pre, 2post, xsbRdata, and xsbWres.  
OrigTotal is the Total time for the original
program from OpenRuleBench, which uses \co{load\_dyn} instead of
\co{load\_dync}.

The results are again as expected.  
For DBLP, XSB is more than three times as fast as Alda, 
9.5 seconds vs.\ 30.6 seconds, but as for PA benchmarks, the overhead of passing the
large data to XSB is large, here 26.9 seconds, and is expected to be reduced to 1\% of that; note that the Alda program has faster reading from
pickled data.

For Wine, XSB is more than eight times as fast as Alda, 3.8 seconds vs.\ 31.0
seconds.  This is due to the use of \co{auto\_table} in Alda generated code,
which does variant tabling, whereas this program, through manual debugging
and optimization, was found to need subsumptive
tabling~\cite{TekLiu11RuleQueryBeat-SIGMOD}.
Optimizations~\cite{LiuSto09Rules-TOPLAS,TekLiu10RuleQuery-PPDP,TekLiu11RuleQueryBeat-SIGMOD}
can be added to the Alda compiler to match this efficiency automatically.
Note that this slowest Alda program is still faster than half of
the systems tested in OpenRuleBench, which took up to 140 seconds and 
three systems gave errors, and where XSB was the fastest at 4.47 seconds~\cite{Lia+09open}.
} 
\noticlp{We examine how the performance scales for large sizes of data and rules
using two benchmarks in OpenRuleBench:
DBLP, the last under large join tests, with the largest real-world data set
among all benchmarks in OpenRuleBench; and Wine, the last under Datalog
recursion, with the largest rule set among all benchmarks in OpenRuleBench.
Again, we changed \co{load\_dyn} used in OpenRuleBench to \co{load\_dync}
for faster reading of facts in XSB's canonical form.

The DBLP benchmark does a 5-way join with projections, 
on DBLP data containing over 2.4 million facts.
The Wine benchmark in OpenRuleBench has 961 rules and 654 facts; it was
originally too slow in XSB but optimized using subsumptive
transformations~\cite{TekLiu11RuleQueryBeat-SIGMOD}, resulting in 967
rules.  The Wine benchmark in Alda is translated from the optimized rules.

Table~\ref{tab-dblp-wine} shows the running times for DBLP and Wine
benchmarks, for both the Alda programs and the XSB programs.
\_extra under Alda is the part of the total time on 2pre, 2post, xsbRdata,
and xsbWres.  OrigTotal under XSB is the Total time for the original
program from OpenRuleBench, which uses \co{load\_dyn} instead of
\co{load\_dync}.  We observe the following results:
} 
\begin{table}[t]
  \small
  \centering
\begin{tabular}{@{\m}l@{\,}||r@{~}|@{~}r||@{~}r@{~}|@{~}r@{~}|@{~}r@{~}|@{~}r@{~}|@{~}r@{~}||@{~}c@{~}||c@{~}|@{\,}r@{\,}}
       & \multicolumn{8}{c||}{Alda} & \multicolumn{2}{c}{XSB}\\
  \cline{2-11} 
  Name & RawR & PickleW 
                      &2pre &xsbRdata &xsbWres &2post &\_extra~&Total &Total 
& OrigTotal \\\noticlp{\hline}\iclp{\cline{1-11}}
  DBLP &12.187 &3.131 &15.722 &11.197 &0.054 &0.020 &26.993 & 30.573 & 9.492
& 63.494\\\noticlp{\hline}\iclp{\cline{1-11}}
  Wine & 0.008 &0.000 & 0.037 & 0.219 &0.000 &0.001 & 0.257 & 30.960 & 3.754
&  3.826\\\noticlp{\hline}\iclp{\cline{1-11}}
\end{tabular}\Vex{.5}
\caption{Running times (in seconds) of DBLP and Wine benchmarks.\Vex{-0}}
\label{tab-dblp-wine}
\end{table}


\noticlp{
\begin{itemize}
\item
For the DBLP benchmark, XSB is more than 3 times as fast as Alda (9.5 vs.\ 30.6 seconds).  The
large data size causes 2pre and xsbRdata (together 26.9 seconds) to dominate the interfacing
overhead (all together 27.0 seconds), as for PA and PAopt benchmarks.
%
Again this overhead of going through files can be reduced to 1\% of it with an
in-memory interface. The remaining time of the Alda program is small (3.6 seconds) because of faster reading from pickled data.

Alda's reading of raw data (12.2 seconds)
is higher than XSB's (7.6 seconds) 
for DBLP
due to the use of Python regular expressions to parse extra string formats
while XSB benefits from drastically reduced checks reading their canonical
data form. This can be fixed with a specialized reading function in C
similar to the one used by XSB.

Note that the original XSB benchmark from OpenRuleBench (63.5 seconds),
without our optimization to use faster data loading, is much slower than
even Alda (30.6 seconds)
that includes the extra overhead.  
OpenRuleBench~\cite{Lia+09open} does not report times for reading data or writing query results, but only query times (0.2--2.4 seconds for DBLP, where XSB was 1.8 seconds, the third fastest, and one system produced an error); the query time in Alda without reading and writing is also small (1.7 seconds).

\item 
For the Wine benchmark,
XSB is more than 8 times as fast as Alda (3.8 vs.\ 31.0 seconds), due to the use of \co{auto\_table}
in Alda generated code.  Manually added subsumptive tabling in the XSB
benchmark from OpenRuleBench reduces the XSB query time from 27.7 
seconds
to 3.7 
seconds.  Again the Alda compiler can be extended to performance such optimizations automatically~\cite{LiuSto09Rules-TOPLAS,TekLiu10RuleQuery-PPDP,TekLiu11RuleQueryBeat-SIGMOD}.

Note that the Alda program is still faster than half of the systems tested in
OpenRuleBench, which took up to 140 seconds and three systems gave errors, and where
XSB was the fastest at 4.47 seconds~\cite{Lia+09open}.
This is despite that each system tested in OpenRuleBench ran the best manually optimized program for it.

\end{itemize}
} 